# Molecular Beam Epitaxy of Ultra-High Quality AlGaAs/GaAs Heterostructures: Enabling Physics in Low-Dimensional Electronic Systems

Michael J. Manfra

Department of Physics, School of Electrical and Computer Engineering, and School of Materials Engineering

Purdue University, West Lafayette IN 47907

Author contact information:

Michael J. Manfra

Purdue University

Email: mmanfra@purdue.edu

Tel: 765-494-3016

**ABSTRACT**

Among very low disorder systems of condensed matter, the high mobility two-dimensional electron gas (2DEG) confined in gallium arsenide (GaAs) – aluminum gallium arsenide (AlGaAs) heterostructures holds a privileged position as platform for the discovery of new electronic states driven by strong Coulomb interactions. Molecular beam epitaxy (MBE), an ultra-high vacuum thin-film deposition technique, produces the highest quality 2DEGs and has played a central role in a number of discoveries that have at their root the interplay of reduced dimensionality, strong electron-electron interactions, and disorder. This review attempts to describe the latest developments in heterostructure design, MBE technology, and our evolving understanding of disorder that result in improved material quality and facilitate discovery of new phenomena at ever finer energy scales.

**INTRODUCTION AND APPROACH**

The importance of innovation in the science and technology of thin film growth in condensed-matter physics cannot be overstated. Development of new, low-disorder, and often highly-engineered materials is central to new discovery in our field. Of the numerous thin-film deposition techniques invented, none has played a more crucial role for the study of low-dimensional electronic systems than molecular beam epitaxy (MBE). MBE's largest impact has been related to the study of the two-dimensional electron gas (2DEG) in AlGaAs/GaAs heterostructures (the discovery of the fractional quantum Hall effect by Tsui, Stormer and Gossard [1] is the most obvious example of unanticipated physics generated in material grown by MBE), but numerous other fields have been fundamentally altered by MBE's ability to epitaxially deposit heterogeneous materials with monolayer precision. Many mesoscopic phenomena in quantum dots and quantum wires have been discovered and fully investigated in materials produced by MBE. MBE and MBE-like techniques have also proven to be useful tools in such diverse areas as high $T_c$ superconductivity [2, 3], growth of complex oxide heterostructures [4, 5], topological insulators [6], and growth of II-VI semiconductor heterostructures culminating in the discovery of quantum spin-Hall effect [7].

This review attempts to describe how MBE is used to produce state-of-the-art low dimensional electronic systems in the AlGaAs/GaAs system needed for experiments designed to explore fundamental phenomena. This review is by no means comprehensive; in fact, we adopt a rather narrow approach. We will not review in any detail the historical development of MBE, it origins dating to the late 1960's in the work of Arthur and Cho at Bell Laboratories [8], nor will we attempt to cover in great detail the many exciting developments in the application of MBE to materials other than GaAs. While application of MBE to other material systems has recently been extremely fruitful, a thorough discussion is left to the experts. We will be brief in our discussion of the features generic to all MBE systems and processes. Many excellent textbooks and reviews describe the fundamental physical mechanisms underlying the MBE growth process as well as the general principles of MBE design and operation [9-14]. The development of the high mobility two-dimensional electron gas in AlGaAs/GaAs heterostructures has a long history which we cannot cover in its entirety [15, 16]. This review will focus instead on recent developments of highly specialized MBE systems dedicated to the production of ultra-clean AlGaAs/GaAs heterostructures with mobility $>10^7 cm^2/Vs$, techniques currently employed by the leading practitioners, and on our evolving understanding of heterostructure design in samples specifically tailored for physics experiments. Much of our discussion will be based on the author's experience in establishing a new ultra-high purity GaAs MBE laboratory at Purdue University in 2011 and lessons learned through several years of collaboration with Loren Pfeiffer and Ken West at Bell Laboratories. We dedicate considerable space to a thorough discussion of the current limits to heterostructure quality, our evolving understanding of disorder in low-dimensional systems and its impact on the physics we hope to explore, and finally, outstanding questions and new directions for research.

**MBE as a tool for physics**

MBE's utility to the physicist derives from its ability to produce extremely clean and structurally-abrupt interfaces of dissimilar materials. This capability in turn allows the MBE grower to control the dimensionality of electronic systems. Indeed, many of the phenomena investigated in materials grown by MBE rely on electronic properties that are unique to the dimensionality imposed by the heterostructure design. MBE has several key attributes that make it ideally suited to creating such structures. At its core, MBE is simply an ultra-high vacuum (UHV) evaporation technique. In an UHV environment, beams of atomic and molecular species are thermally evaporated and are incorporated into a heated substrate placed in the line-of-sight of the emerging beams. The chemical composition of the growing film is controlled by mechanical shutters placed in front of the thermal beams. Distinguishing features of MBE include:

1) Large area, single-crystal films with extremely low extended defect density and exact epitaxial registry between the starting substrate and the overgrown film can be produced
2) Unintentional impurity incorporation can be extremely low – on the order of $10^{13}$/cm$^3$ in the best GaAs MBE systems
3) MBE is a slow growth technique, typical growth rates are between 0.1 and 2 monolayers/sec – at this rate atomically abrupt interfaces can be produced
4) UHV conditions allow numerous in-situ diagnostic techniques, giving the grower real-time feedback on the state of the growing film

A number of inherent properties make AlGaAs/GaAs a model system for both MBE growth and exploration of low-dimensional electron physics. Figure 1 displays a scanning transmission electron microscope image of a 7nm AlAs/5nm GaAs 50 period superlattice. The bright strips are the GaAs layers while the darker strips are AlAs. The sharpness of the heterointerfaces and the coherency of the superlattice are evident. The lattice mismatch between AlAs and GaAs is only 0.1%, allowing full range of alloy composition and layer thicknesses to be grown without the formation of extended defects. The AlGaAs/GaAs heterostructure system has a type I band alignment; the AlGaAs barrier acts as a barrier for both conduction band electrons and valence band holes and can provide upwards of 0.3eV of confinement for electrons at an $Al_{0.35}Ga_{0.65}As$/GaAs interface. The effective mass of conduction band electrons in GaAs is also relatively light, m*=0.067$m_e$, where $m_e$ is the free electron mass. The light mass results in high electron mobility and strong electrostatic confinement in heterostructures. There are also a number of practical considerations worth mentioning. Driven by years of research and industrial demand, high quality bulk GaAs substrates are readily available and fairly inexpensive. Furthermore, starting materials including arsenic, gallium, and aluminum can be obtained from a number of producers and purification techniques for these elements are sufficiently well-developed to enable high purity MBE growth of their compounds. These last two points are important starting conditions for a sustained effort at improving heterostructure quality via MBE growth to the level obtained in GaAs. The importance of a well-developed substrate technology for GaAs can be contrasted with the situation in another III-V semiconductor system, AlGaN/GaN, in which lack of widely-available low defect density GaN substrates presents a substantial impediment to progress [17-19].

**The two-dimensional electron gas and mobility**

The workhorse of low-dimensional electron physics is the two-dimensional electron gas. Not only does examination of the properties of 2DEG itself remains an active area of research but the 2DEG also forms a basic building block for a number of other low-dimensional systems, e.g. quantum dots and quantum wires. For 30 years, low-temperature mobility has been used to benchmark 2DEG quality in the AlGaAs/GaAs system. Mobility is defined as $\mu=\sigma/en_e$ where $\sigma$ is the conductivity, e is the charge of the electron and $n_e$ is the 2D areal density of electrons. Crudely speaking, mobility measures an electron's ability to carry current without undergoing large-angle scattering – higher mobility implies less large-angle scattering. Mobility has improved from 5,000$cm^2$/Vs for early modulation-doped samples in 1979 [20, 21] to over $3x10^7 cm^2/Vs$ in present-day state-of-the-art samples [22, 23], an amazing improvement. To put this in context; an electron in a $3x10^7 cm^2/Vs$ sample has at low temperatures a ballistic mean free path of 0.3 mm, essentially traveling macroscopic distances between hard scattering events! Low temperature 2DEG mobility is a useful metric for the MBE grower for several reasons. Even in the highest quality AlGaAs/GaAs 2DEGs currently available, mobility tends to saturate at temperatures below T=1K. At T=1K, acoustic phonon scattering is sufficiently weak such that it can be neglected [24]; thus any saturation of mobility must be associated with static and largely temperature independent scattering centers. Therefore mobility can be used to quantify the "residual disorder" in the MBE-grown sample in a fairly unambiguous manner (we will discuss the utility of mobility as an indicator of high magnetic field transport in a later section). Scattering from residual disorder comes in several flavors; remote charged impurity scattering from intentional silicon donors, alloy disorder scattering, interface roughness scattering, and uniformly distributed background charged impurities are a few examples. Importantly to the MBE grower, mobility is also easy to measure. It requires minimal processing (employing the van der Pauw geometry) and only a simultaneous measurement of resistivity and the low magnetic field Hall effect is necessary to characterize a newly grown wafer. The importance of this fact cannot be overestimated. One of the principal challenges of ultra-high purity MBE growth is shortening the feedback loop between sample growth and electrical characterization. The MBE grower needs some means to determine if a variation in a growth parameter has a positive or detrimental impact on 2DEG quality. While it is not the only parameter important for 2DEG functionality, low temperature mobility is the quickest means to assess sample quality. The humble mobility measurement has become an invaluable tool for the MBE grower as he/she tries to efficiently optimize growth conditions and heterostructure design. In the author's laboratory at Purdue, two wafers are grown per day, and as needed, these two growths can be fully characterized at T=0.3K and high magnetic field the next day, providing valuable input for subsequent growths.

**MBE SYSTEM CONSIDERATIONS**

Given the correlation between sample quality and visibility of low-energy scale physics, it not surprising that several research groups around the world have dedicated considerable effort to developing MBE growth systems focused primarily on the achievement of ultra-high mobility 2DEGs in GaAs. The Purdue group, the Princeton/Bell Labs group, and the Weizmann group have all reported 2DEG mobility in excess of $20 \times 10^6 cm^2/Vs$. The samples of all three of these groups are used extensively to study the fragile fractional quantum Hall states of the 2$^{nd}$ Landau level, one of the most exciting topics in contemporary condensed-matter research. The Princeton/Bell Labs group and the Weizmann group have both reported peak mobility above $30 \times 10^6 cm^2/Vs$ [22, 23]. Investigation of the 2$^{nd}$ Landau level requires samples of the highest quality. In order to work in this regime, AlGaAs/GaAs MBE hardware has become highly specialized. Several factors influence the design of MBE systems dedicated high-mobility AlGaAs/GaAs growth. Vacuum pumping speed and ultimate base pressure, thermal efficiency of heaters, cleanliness of materials subjected to high temperatures, subsystem redundancy, and safeguards against common system failures are principal concerns. The MBE system in use at Purdue is shown in Figure 2. Working closely with the author, this highly customized system was built by Veeco Inc. [25]. The design is based on the venerable GenII MBE, a system originally designed and built by Varian Associates starting in the 1980's. However, most of the critical components, as well as fundamental aspects of the growth chamber's geometry, have been modified to a significant extent in the Purdue version.

Since charged impurities in the GaAs lattice are a significant source of mobility-limiting scattering, their elimination is of paramount importance. Unintentional charged impurities can come from a number of sources including the background vacuum, hot metal surfaces within the growth chamber, the starting GaAs substrate, and the starting elemental materials used for semiconductor growth. We begin by considering the vacuum system. As shown in Figure 1, the Purdue system consists of three separate chambers: a loading chamber, a buffer/outgassing chamber and the main growth chamber. All three chambers are pumped by closed-cycle helium cryopumps with base pressures of $\sim 10^{-10}$ torr, $\sim 10^{-11}$ torr and $\sim 10^{-12}$ torr respectively. In total, there are five closed-cycle helium cryopumps on this system. The growth chamber alone is pumped by three 3000 liter/sec closed-cycle helium cryopumps, a liquid nitrogen-cooled titanium sublimation pump and a liquid-nitrogen-filled panel surrounding the sample and source furnaces within the vacuum space. As was first demonstrated by Pfeiffer and West, the entire system, including the external walls of the cryopumps, is baked at 200°C for extended periods of time [26]. For this scheme to work, all components in the MBE growth chamber must have exceedingly low vapor pressures at temperatures exceeding 200°C and must be robust against damage during extended bake-outs. Surprisingly enough, this precludes a number of items commonly found in standard commercial MBE systems, including, for example, elastomeric seals found in most gate valves as well as some high vapor pressure (at high temperature) materials sometimes used in conjunction with moving parts within the vacuum and in the cryopumps themselves. During the installation of the Purdue system in late 2010, the MBE was baked for a total of six weeks. Following bake-out, the internal liquid-nitrogen-filled cyropanel is cooled to T=77K and is maintained in this condition for the remainder of the growth campaign, which can last several years with proper planning and good fortune. The combination of extremely high pumping speed, meticulous choice of UHV- and high temperature-compatible materials, and extensive baking results in MBE base pressures as low as $1 \times 10^{-12}$ torr. Note that to measure pressures below a value of $\sim 2 \times 10^{-11}$ torr, the ion gauge and electronics must be of a special design [27]. At $1 \times 10^{-12}$ torr the background pressure in the MBE is dominated by $H_2$ molecules emanating from the stainless steel walls of the chamber. The partial pressures of all other atomic and molecular species are at least 1 to 2 orders of magnitude lower and most are below the detection limit of even the most sensitive commercially available residual gas analyzers (RGA) [28]. Hydrogen, it turns out, does not seem to adversely affect the quality of GaAs epilayers and may be in fact beneficial [29, 30].

Achievement of deep UHV conditions in the growth chamber is an important first step to high purity GaAs growth, but other sources of impurities are equally important. Consider for example the components that get hot during normal MBE operation. During MBE growth the effusion cells (source material furnaces) must maintain the elemental gallium, aluminum and arsenic at elevated temperatures to create molecular beams. Arsenic is a high vapor pressure material so raising its temperature to 350°C is sufficient for GaAs growth. However, the arsenic cell is large to accommodate a 2.5kg charge and under typical operating conditions it dissipates approximately 150 Watts. Examination of the equilibrium vapor pressure vs. temperature profiles for gallium and aluminum highlights another challenge. The vapor pressures of aluminum and gallium only exceed $10^{-5}$ torr above 900°C and 800°C respectively. For a typical AlGaAs/GaAs growth in our chamber the Ga cell is held at 850°C and the Al cell is at 980°C as measured by a thermocouple in contact with a pyrolytic boron-nitride crucible containing the molten metal. The actual heating filaments are undoubtedly hotter still. While the aluminum and gallium cells are smaller in size than the arsenic cell, they still present a significant thermal load on the MBE (each dissipates about 150 Watts) and the hot surfaces of the effusion cell are a potential source of contamination. The effusion cells can also radiatively heat other surfaces within the MBE causing additional outgassing. Thus the cleanliness and thermal efficiency of the cell design become important considerations for high mobility growth. For ultra-high purity GaAs systems, custom-designed cells with high density heating elements and enhanced radiation shielding are employed. All of the effusion cells in the Purdue system have been significantly modified beyond standard vendor products. Meticulous handling and cleaning procedures for these cells must also be used. It is routine to outgas the cells near 1600°C in an ancillary UHV chamber prior to loading into the MBE.

The operation and design of the substrate manipulator and heater must also be considered. During growth, the GaAs substrate is normally maintained at 635°C, near the congruent sublimation temperature of GaAs [31]. In the Purdue system, the substrate is mounted on a custom-designed puck made of high purity tantalum using liquid gallium metal as "glue". The gallium metal holds the GaAs substrate in place by surface tension and provides good thermal contact to the tantalum puck. The tantalum puck is radiatively heated by filaments mounted in the manipulator. In order to heat the substrate to 635°C, the heater consumes approximately 150 Watts. The heater can outgas impurities that eventually find their way into the growing GaAs lattice. In 1997, Umansky *et al*. [32] reported a significant improvement in 2DEG low-temperature mobility ($8 \times 10^6 cm^2/Vs$ to $14 \times 10^6 cm^2/Vs$) when they switched from a 3 inch substrate heater to a 2 inch version that consumed 30% less power. Moreover, the substrate is typically rotated at 10 rpm in order to maintain good growth uniformity across the wafer. Moving parts within an UHV environment act as another potential source of impurities and methods of lubrication that might be acceptable in a less demanding MBE application must be evaluated for their potential impact on GaAs quality.

MBE systems used for ultra-high purity GaAs growth must also be able sustain long, uninterrupted growth campaigns. Once the system is producing high-mobility material, any loss of the vacuum integrity will inevitably result a significant decline in material quality. Depending on the severity of system compromise, it can take several months of constant growing to recover to peak mobility values. In the Purdue laboratory all critical electrical systems are on an uninterruptable power supply. This includes not only computers, system control electronics, and effusion cell power supplies, but also the four 5000 Watt compressors responsible for running the 5 cryopumps on the system. The liquid nitrogen supplied to the internal panels within the MBE must also be maintained without interruption. To buffer our system against possible failures of the house nitrogen delivery system, we have installed a 1000 liter "microbulk" storage tank immediately adjacent to our MBE that is constantly full. If the house nitrogen delivery system fails for any reason, our MBE is programmed to switch over to the microbulk tank. This local tank is sized to last approximately 3 days of continuous operation until it needs to be refilled. The expectation is that the main delivery system can be restored within this time period. The effusion cells are also loaded redundantly. Silicon is used as the n-type dopant while carbon is used to produce p-type material. Both of these are filament sources with two independent filaments per source. If either the primary Si or C filament breaks, another is available. Typically two or three cells are filled with gallium, two cells are filled with aluminum and approximately 2.5 kilograms of arsenic is loaded into a very large cell. This configuration of materials not only allows for construction of multiple alloy composition materials in a single growth, but using three gallium cells and two aluminum cells also prolongs the campaign and provides a safeguard if one of effusion cells should fail.

While not strictly-speaking a machine design consideration, we will discuss shortly the importance of proper choice and handling of source material used for ultra-high purity MBE growth. At present, it appears that impurities originating in the source gallium, aluminum and arsenic are responsible for limits on low-temperature 2DEG mobility.

**WHAT LIMITS MOBILITY IN STATE-OF-THE-ART HETEROSTRUCTURES?**

Low-temperature mobility is used as a standard metric of 2DEG quality. It is natural to ask what limits mobility in the best material. It has been known for quite some time that in the regime of large delta-doping setbacks, uniformly distributed charged impurities in the vicinity of the 2DEG are the principle source of scattering [26, 32] in very high mobility material. The majority of this scattering seems not to originate from the ionized silicon impurities introduced during modulation-doping, but rather, from unintentionally incorporated impurities uniformly distributed in the AlGaAs/GaAs lattice. This conclusion is supported by wealth of experimental and theoretical evidence and can be seen most clearly in the dependence of mobility on 2DEG density. Calculations (see Ref. [24] and references therein) indicate that for mobility limited by remote ionized silicon donors, $\mu \sim n_e^\alpha$ with $\alpha \sim 1.5$ is expected. For scattering dominated by uniformly distributed background impurities, $\alpha$ is approximately 0.7. Numerous experiments using large doping setbacks yield $\alpha$ consistent with mobility limited by uniformly distributed background impurity scattering [26, 32]. Representative early data from the Bell Labs group and the Weizmann group with mobility around $10^7 \text{cm}^2/\text{Vs}$ is shown in Figure 3. It is important to note that the data of Figure 3 were obtained utilizing a single heterojunction design, typically a single $Al_{0.35}Ga_{0.65}As$/GaAs interface, in which the silicon dopant atoms were placed at a setback of at least 50nm. We will discuss the impact of using quantum wells instead of single interfaces shortly.

Given the data of Figure 3, several questions immediately follow. How many uniformly distributed background charged impurities are in the vicinity of the AlGaAs/GaAs interface? What impurity atoms are most prevalent and where do they come from? These questions are difficult to answer directly with standard spectroscopy techniques such as photoluminescence [26], secondary ion mass spectrometry [33], or deep level transient spectroscopy [34, 35]. With total impurity concentrations significantly below $1x10^{14} cm^{-3}$ in state-of-the-art material, these traditional characterization techniques are not sufficiently sensitive. Most of what we know is inferred from transport studies. From calculations of 2DEG mobility dependence on background impurity concentration we can estimate that the total density of ionized impurities in the best material today is approximately $5x10^{13}/cm^3$ [24]. This of course does not tell us the atomic species or where they came from, but the calculations do suggest that if the MBE grower can somehow reduce the total background charged impurity concentration down to approximately $10^{12}/cm^3$, low temperature mobility could approach $100x10^6 cm^2/Vs$ [24]. The calculation of Hwang and Das Sarma displayed in Figure 4 indicates the expected dependence of mobility on background charged impurity density.

To improve sample quality the MBE grower has two options: reduce the background impurity concentration or utilize heterostructure designs that render the remaining impurity scattering less significant. We first consider reduction of background impurities. As mentioned, direct spectroscopic probing of the remaining uniformly distributed background impurities in the highest purity GaAs is difficult. Historically, however, several mobility-limiting impurities found in MBE-grown GaAs and AlGaAs have been identified including carbon, oxygen, sulfur and silicon [36-39]. It is reasonable to assume that the same impurities are present in the highest mobility AlGaAs/GaAs 2DEGs, just at substantially reduced levels.

Evidence suggests that the background vacuum is not the dominant source of contamination in the best systems. Figure 5 displays a spectrum of residual gases found in the Purdue system early in its initial growth campaign, the day after the growth of an AlGaAs/GaAs heterostructure with mobility greater than $10^7 cm^2/Vs$. The dominant species, other than omnipresent hydrogen, is arsenic as expected. Quantitative analysis of other residual species in this UHV regime is difficult and it should be noted that the filament of the residual gas analyzer itself is known to produce carbon monoxide and carbon dioxide (amu/e 28 and 44 respectively) further complicating quantitative analysis [40] in the deep UHV regime. Nevertheless, this spectrum indicates that the growth process does not significantly corrupt the original vacuum as the vacuum is as clean, or even cleaner, than that observed immediately after system bake-out. Of course the residual gas analyzer has a sensitively limit of approximately $10^{-14}$ torr so absence of impurities in a residual gas spectrum does not guarantee that grown GaAs is impurity free. More substantial evidence that the vacuum quality is not the principal source of impurities in the current generation of samples comes from growth experiments in which a growth pause is inserted exactly at the heterointerface where the 2DEG will reside. Located exactly at the position of the 2DEG, any impurities accumulated from the vacuum during this pause will prove highly detrimental to mobility. Pauses ranging from several seconds to several minutes at the interface had no statistically significant impact on our measured mobility. In fact, 20 second pauses are standard for each heterointerface forming the quantum well in our structures with mobility above $20x10^6 cm^2/Vs$.

The nature and relative significance for high mobility growth of impurities emanating from the aluminum, gallium, and arsenic source material is not a resolved issue. High purity aluminum is of course crucial for growth of high quality AlGaAs, the material used as a barrier for 2DEG structures. In addition to impurities emanating directly from the aluminum source, aluminum is extremely reactive and is a known getter of oxygen. Oxygen forms a deep level in GaAs and AlGaAs [14]. Sulfur and carbon are known to be significant impurities in arsenic. As sulfur and carbon are acceptors in GaAs, the purity of the arsenic charge used in MBE growth has received considerable attention. Using DLTS and multiple arsenic sources during a single campaign, Chand *et al.* [37, 38] concluded that arsenic is the principal source of residual acceptor impurities in GaAs. Similar conclusions were drawn by Umansky *et al.* for high mobility AlGaAs heterostructures; they reported a gradual improvement in 2DEG mobility as their arsenic charge was depleted [32].

Gallium purity has received less attention. Schmult and collaborators [41] recently performed an analysis of compounds generated during an initial outgassing of a new gallium charge. They determined that the principal impurities associated with a new gallium source are $GaO_2$ and $GaH_3O$, at least in the sensitivity range of their residual gas analyzer (~$10^{-13}$ torr partial pressure). These signals were seen to diminish within a few thermal cycles of the gallium to growth temperature. Of course the sensitively of any current residual gas analyzer is insufficient to detect all impurities that may limit mobility above $10^7 cm^2/Vs$. Data from the initial growth campaign at Purdue suggest that in fact that the gallium source material, and to a lesser extent the aluminum source material, but not the currently used arsenic source, is the dominant source of residual acceptor impurities in our system. This conclusion is based on the following observations during the early growths in our new machine. Despite achieving a background vacuum of ~1x10$^{-12}$ torr and meticulous handling and treatment of our aluminum, gallium, and arsenic source material, our initial attempts to grow a 2DEG in a single heterojunction (SHJ) Al$_{0.33}$Ga$_{0.67}$As/GaAs structure were unsuccessful. It was determined that undoped bulk GaAs grown in our system exhibited an extremely high p-type background conductivity (p>>10$^{14}$cm$^{-3}$ in the earliest test structures). This p-type background, generated by unintentional acceptors, was only removed by vigorous high temperature outgassing of the gallium source material. In order to achieve our current mobility $> 20 \times 10^6$cm$^2$/Vs, the gallium was subjected to several bakes in which the cell was run *at 200°C above normal growth conditions for several hours*. While this procedure certainly wasted a significant quantity of usable gallium metal, its purpose was to preferentially drive off impurities with higher vapor pressure than gallium. Each thermal treatment resulted in a measurable improvement in material quality. During this series of outgassing experiments, the arsenic was not given any treatment; the gains in mobility can be directly attributed to improvement in gallium purity. We also outgassed our aluminum source significantly above growth temperature, but this resulted in less significant gains. It is also worth noting that since the electron's wavefunction resides primarily in GaAs, not the AlGaAs barrier, mobility will be most sensitive to the quality of the GaAs channel. This experience with source conditioning yielded valuable lessons. Source material purity is most likely the factor limiting further improvement in 2DEG mobility. Importantly, source material quality can be improved substantially *in-situ* during MBE operation. Both facts suggest routes to further improvements in 2DEG quality that are currently under investigation in our laboratory.

**Heterostructure design considerations**

So far we have paid scant attention to heterostructure design. In recent years it has become evident that heterostructure design is just as important as starting material purity to the attainment of ultra-high quality 2DEGs. Moreover, evidence is now accumulating that heterostructure design and the specific nature of disorder present in a given sample dictate the strength of fragile fractional quantum Hall states to an extent that has previously not been appreciated. It is important to note these details are not fully characterized by simple mobility measurements at zero magnetic field. The relationship between mobility, heterostructure design, and visibility of exotic correlated states in the 2$^{nd}$ Landau level will be discussed shortly. We first discuss the relationship between heterostructure design and mobility.

The simplest heterostructure design to produce a high mobility 2DEG is the single heterojunction. Consider an $Al_{0.35}Ga_{0.65}As$ barrier is deposited on a thick (~1μm) layer of GaAs. The barrier is 240nm thick and the entire structure is capped with 10nm of GaAs. This particular structure is delta-doped with silicon (density $8x10^{11}cm^{-2}$) in the AlGaAs layer at a setback of 70nm from the GaAs channel layer. The silicon doping is responsible for band bending and transfers charge to the heterojunction. The conduction band edge as a function of position and the resulting free charge density at the heterointerface is shown in Figure 6. This SHJ will typically produce a 2DEG density of $2.2-2.4x10^{11}cm^{-2}$ after illumination with red light at low temperatures. Illumination is needed since in $Al_xGa_{1-x}As$ with $x > 0.20$, the silicon donor is no longer shallow ($E_a$~10meV) and forms a deep donor state ($E_a$~135meV), the so-called DX center [42-44]. Illumination facilitates charge transfer to the 2DEG at the heterojunction. While the SHJ is heavily utilized in experiments to this day, the design has limitations. As is evident from the mobility vs. density data of Figure 3, mobility tends to increase as the 2DEG density is increased. Naively we can expect higher mobility if the 2DEG density can be increased beyond ~$2.2 \times 10^{11}$cm$^{-2}$ in the SHJ. In practice however this is difficult. In order to increase the 2DEG density we must decrease the delta-doping setback; this in turn increases the scattering from the ionized donor silicon atoms. Higher density also forces the 2DEG closer to the AlGaAs/GaAs heterointerface, enhancing interface roughness scattering. While higher density can be achieved in the SHJ, mobility is not enhanced.

An obvious way to overcome this limitation is to build a quantum well with AlGaAs barriers on both sides of a thin GaAs channel layer. The AlGaAs barriers can then be symmetrically doped with silicon from both sides such that large setbacks can be maintained while increasing the 2DEG density significantly above $2.2 \times 10^{11}$cm$^{-2}$. The highest mobility 2DEGs grown today are all quantum well structures. In reality the highest mobility heterostructures, the ones also typically used to study fragile quantum Hall states such as $\nu=5/2$ and $\nu=12/5$ in the 2$^{nd}$ Landau level, are more complicated than what has just been described and involve one of several variations of short-period superlattice doping. The short-period superlattice scheme was first introduced by Friedland *et al*. [45] and later discussed by Umansky *et al*. [23]. The conduction band edge profile of a modern quantum well design is shown in Figure 7a. (To the best of our knowledge this design was first utilized in the context of ultra-high mobility growth by the Bell Labs group, e.g. Ref. [22].) Several key differences from the SHJ are immediately evident. The 2DEG is confined in a 30nm GaAs quantum well bounded by $Al_{0.24}Ga_{0.76}As$ barriers. The lower aluminum content of the barrier is still sufficient for electron confinement while reducing interface roughness and incorporation of unintentional impurities associated with higher mole fraction barriers. It is important to note that the silicon dopant atoms are not placed directly in the AlGaAs barrier, but rather, are placed in extremely narrow (3nm) GaAs "doping-wells" surrounded by 2nm pure AlAs barriers. The doping level is usually quite high; for structures grown at Purdue $1 \times 10^{12} cm^{-2}$ silicon atoms are placed in the doping-well above the main quantum well and $0.8 \times 10^{12} cm^{-2}$ are placed in the doping-well located on the substrate side of the main well. This has important consequences; a detailed view of the band structure around the lower doping-well is shown in Fig. 7b. Note that only $3 \times 10^{11} cm^{-2}$ carriers are ultimately transferred to the 2DEG in the principal quantum well. While some of the charge from the upper doping-well is transferred to the sample surface to compensate surface states, a significant amount of charge remains that is transferred neither to the sample surface nor to the primary 2DEG. Our simulations indicate that charge moves from the silicon parent atoms to the thin AlAs barriers surrounding the doping-well, as indicated in Fig. 7b. While the AlAs acts as a barrier at the $\Gamma$ point as indicated in the Figure 7, for x above ~0.40 the $\Gamma$ and X bands in $Al_xGa_{1-x}As$ cross, so that the X point band edge is actually below the GaAs $\Gamma$ point. Interestingly, there is a range of doping in which the excess charge transferred to the AlAs barriers does not appear as a parallel conduction path in low frequency transport at low temperatures and high magnetic fields. In Figure 8 we show magnetotransport at T=0.3K from such a design grown at Purdue. The absence of significant parallel conduction is evidenced by the zeroes of longitudinal resistance in the quantum Hall regime. Yet these excess carriers still play an important role, presumably screening the potential of the parent ions. Since the silicon atoms are placed in GaAs, not AlGaAs, there are no associated DX centers. Silicon incorporates as a shallow ($E_a$~5meV) center in GaAs. The transfer of charge from the narrow GaAs doping-well to the primary quantum well is driven by the difference in confinement energies. Thus no illumination is required to achieve maximum 2DEG density, although low temperature illumination is still often used to improve the quality of the fractional quantum Hall states. The doping-well design not only provides the highest mobility but has been a boon for exploration of exotic and small-gapped fraction quantum Hall states, but it must be stated that the design also has its limitations associated with charge instabilities near the doping-well. These issues will be discussed shortly. In addition the doping-well design is also helping to reshape our understanding of the role of disorder and the utility of mobility as a metric of 2DEG quality.

**SAMPLE DESIGN CONSIDERATIONS FOR THE FRACTIONAL QUANTUM HALL REGIME**

**How predictive is zero-field mobility as an indicator of high magnetic field correlations?**

Throughout this review we have stressed the importance of low temperature mobility as a metric of 2DEG quality. Yet quantitative incorporation of disorder effects into any theory of the fractional quantum Hall effect is notoriously difficult and the experimental correlation between zero-field mobility and excitation gaps in the fractional quantum Hall effect (FQHE) has never been very strong. While is it clear that mobility is a useful quantity for initial sample screening, its connection to physics at high magnetic fields deserves careful consideration. It is understood that the visibility of the FQHE depends on a subtle interplay of disorder and strong electron-electron interactions. Whether the disorder most relevant for the FQHE is best quantified by zero-field mobility is the question we now address.

**Fractional quantum Hall physics in the second Landau level**

The fractional quantum Hall effect is characterized by vanishing longitudinal resistance and quantized Hall resistance at specific rational fractional values of filling factor $\nu$. Filling factor ($\nu$) is defined as the ratio of the areal electron density to the density of magnetic flux quanta and corresponds to the number of filled Landau levels. Integral filling factor corresponds to the Fermi level residing in a gap between a completely filled Landau level and a completely empty level; fractional filling indicates a partially filled Landau level without an associated gap in the single-particle density of states. The fractional quantum Hall effect occurs at certain magic values of rational fractional filling (e.g. 1/3, 1/5, 2/3 …) where strong electron-electron interactions introduce a gap to low-lying excitations. The extremely fragile fractional quantum Hall states in the $2^{nd}$ Landau level are presently the subject of intense scrutiny, especially the even-denominator state at $\nu=5/2$. The $\nu=5/2$ state does not obey the normal odd-denominator rule exhibited by all single layer states in the lowest Landau level and thus cannot be described by a hierarchical Laughlin-like wavefunction [46-50]. It is now widely believed, but not conclusively proven, that the $\nu=5/2$ ground-state is described by the so-called Moore-Read Pfaffian wave function [51] or some closely related state such as the anti-Pfaffian [52]. Crudely speaking, the existence of a gapped state at $\nu=5/2$ is ascribed to a p-wave BCS-like pairing of composite fermions [51, 53-56]. If this description is indeed true, several important consequences follow. The low-lying charged excitations of the Pfaffian state are believed to possess non-Abelian braiding statistics [51, 55, 57-62]. For particles obeying non-Abelian statistics repeated interchange of two identical particles does not change the many-body wavefunction by a factor of +/-1 as for bosons and fermions, respectively, but rather, produces a unitary transformation of the wavefunction within a degenerate manifold. The existence of excitations with non-Abelian statistics potentially has implications for quantum computing [57-62] as computations with non-Abelian excitations would be topologically protected against decoherence. Decoherence is a major obstacle for implementation of all known solid-state quantum computing platforms. This exciting combination of new physics and potential for applications has driven a world-wide experimental effort to understand the nature of the $\nu=5/2$ state.

**Sample quality and the $\nu=5/2$ state**

The $\nu=5/2$ state was first observed by Willett *et al*. [63] in a single heterojunction sample with mobility $1.3 \times 10^6 \mathrm{cm}^2/\mathrm{Vs}$. However the state was extremely weak in these early samples; the longitudinal resistance was not activated as expected for a gapped state and the Hall resistance was not well-quantized. True quantization and activated transport was first reported by Eisenstein *et al*. [64] in a sample with mobility $7 \times 10^6 \mathrm{cm}^2/\mathrm{Vs}$ and later confirmed in Ref. [65] in a $17 \times 10^6 \mathrm{cm}^2/\mathrm{Vs}$ mobility sample. At the time it was assumed that as 2DEG mobility improved, the transport features would improve as well. To some degree this was true, the original observation of Willett *et al*. did not show activated transport at $\nu=5/2$; while later, higher mobility, samples did. Close inspection however reveals a more complicated situation. In 1990, in the $7 \times 10^6 \mathrm{cm}^2/\mathrm{Vs}$ sample, the energy gap at $\nu=5/2$ was measured via activated transport to be $\Delta_{5/2}=105 \mathrm{mK}$; in 1999 in the $17 \times 10^6 \mathrm{cm}^2/\mathrm{Vs}$ mobility sample the gap was still only 110mK. Note that both samples had the exact same density and the same heterostructure design, making a comparison meaningful. A greater than factor of 2 increase in mobility did not improve the strength of the gap at $\nu=5/2$. Moreover both measurements are significantly below theoretical estimates of 1-2 K [66-68] expected for $\nu=5/2$ in the absence of disorder.

A significant improvement in transport in the 2nd Landau level was associated with the use of doping-well samples produced at Bell Laboratories. Using a sample with $n=3 \times 10^{11} \mathrm{cm}^{-2}$ and $\mu=31 \times 10^6 \mathrm{cm}^2/\mathrm{Vs}$, Eisenstein *et al*. [22] reported a 5/2 gap $\Delta \approx 300 \mathrm{mK}$ and the first systematic identification of the reentrant integer quantum Hall effect for $2 \leq \nu \leq 4$. Using another piece from the same wafer, Xia *et al*. [69] reported on several new features in the 2$^{nd}$ Landau level observed at ultra-low temperatures below T=10mK including the first observation of a FQHE at ν=12/5, another possible non-Abelian state [70]. In a later set of experiments again using the same material, Pan *et al*. reported an excitation gap at ν=5/2 of approximately 450mK [71]. More recently Kumar *et al*. [72] reported observation of a new fractional state in the 2$^{nd}$ Landau level at ν=2+6/13. This sample was also used to study the collective nature of the reentrant integer quantum Hall states in the second Landau level [72]. While these doping well samples certainly have higher mobility than the SHJ samples, the extremely limited number of data sets preclude any strong conclusions correlating mobility and strength of the 5/2 gap.

The dependence of the energy gap at ν=5/2 on mobility was studied by three groups in 2008 [71, 73, 74]. Dean *et al*. [74] studied the excitation gap $\Delta^{norm}=\Delta_{5/2}/(e^2/\varepsilon l)$, normalized to the strength of the electron-electron interaction $e^2/\varepsilon l$, where e is the charge of the electron, ε=12.9 is the GaAs dielectric constant and $l = \sqrt{\hbar/eB}$ is the magnetic length, vs. the inverse transport lifetime deduced from mobility for their low density sample as well as for other samples reported in the literature in an attempt to extrapolate to a disorder-free intrinsic gap value. A significant difference between the extrapolated disorder-free gap and theoretical estimates was noted. To quantify the impact of disorder, Pan *et al*. [71] plotted the normalized energy gap at ν=5/2 vs. inverse mobility for several high mobility samples including data taken from the literature. These samples were of different designs and 2DEG density, nevertheless a general trend of decreasing energy gap with decreasing mobility was claimed; their data is reproduced in Figure 9. Despite the substantial scatter in the data, the dashed line in Figure 9 is taken as a fit and implied a "mobility threshold" of approximately $10^7$cm$^2$/Vs for observation of a gapped state at ν=5/2.

An important step in understanding what type of disorder might be most relevant for the 2$^{nd}$ Landau level was taken in a combined experiment and theory work reported by Nuebler *et al.* [75]. In a back-gated sample Nuebler studied the density dependence of the energy gap at 5/2 filling. They reported a gap $\Delta_{5/2}$=310mK at the highest density measured. They noted the lack of correspondence between the actual measured gap strength and that naively expected from theory. They also plotted the measured gaps at ν=5/2 from their variable density sample along with other values reported in the literature, finding little correspondence between gap strength and mobility. This data is displayed in Figure 10. Based on this data, they noted that mobility is a poor figure of merit to predict the quality of the ν=5/2 state. Nuebler *et al.* [75] also reported calculations including finite well-width and Landau level mixing effects. An important conclusion of this work was that the excitation gap should be strongly influenced by the disorder induced by the remote silicon donor impurities; this assertion was based on the size of the ν=5/2 quasiparticles determined from numerical calculations. They estimated the excitations at 5/2 are at least 12 magnetic lengths in diameter - approximately 150nm at a magnetic field of 4T. This length scale is comparable to the setback to the silicon donors used in modern heterostructure designs.

Data from the Purdue group also suggests that mobility is a poor metric to quantify disorder relevant to 5/2 physics. During our initial growth campaign we tracked the improvement of mobility with growth number and changes to structural design details. We also measured all of our samples at T=0.3K and high magnetic fields to assess the quality of the developing fractional quantum Hall states. The results obtained with doping-well samples were somewhat surprising. An example for an early doping-well sample with mobility of only 11x10$^6$cm$^2$/Vs is shown in Figure 11. Despite the low mobility and elevated temperature, many nascent features in the 2$^{nd}$ and higher Landau levels are already discerned. In particular, the resistance $R_{xx}$ at ν=5/2 is quite low, ~25Ω, and very symmetric about half-filling. Crudely speaking, one might say that the quality of the fractions looks "better" than one might naively expect for this mobility. In an attempt to quantify this behavior at T=0.3K we began characterizing our samples not only by zero-field mobility but also by a "resistivity at ν=5/2". For this measurement, we simply repeat the determination of resistivity conducted at zero magnetic field in our van der Pauw geometry, but now with the magnetic field tuned to exactly 5/2 filling. For example the resistivity at ν=5/2 ($ρ_{5/2}$) for the sample in Figure 11 was 40.8Ω/□. This sample was further cooled to ultra-low temperatures with surprising results. Figure 12 is taken from the work of Samkharadze *et al*. [76] in which the same sample was cooled to T~5mK. All major fractions of the 2$^{nd}$ Landau level are well-developed, including the elusive ν=12/5 state. The 5/2 excitation gap is $Δ_{5/2}$=450mK, among the highest ever measured, despite the fact that the mobility is only 11 million – a value heretofore considered on the boundary for a gapped state at 5/2 [71]. Based on these results, the Purdue group now uses the smallness of the resistivity at ν=5/2 at T=0.3K as the primary indicator of 2DEG quality in doping well samples. The lowest $ρ_{5/2}$ we typically observe at 2DEG density ~3x10$^{11}$cm$^{-2}$ is around 30Ω/□ in the doping-well design. It is typically higher with other heterostructures designs (e.g. doping directly in the AlGaAs barrier), even in samples in which the mobility exceeds 20x10$^6$cm$^2$/Vs.

Further evidence questioning the correlation of mobility and physics in the 2$^{nd}$ Landau level which also supports use of $ρ_{5/2}$ as useful metric of 2DEG quality was found in later Purdue samples. As we continued to grow wafers, mobility unsurprisingly continued to improve. However, we based our decisions about which samples to cool to lowest temperatures based on the behavior exhibited around ν=5/2 at T=0.3K. Figure 13 shows and overview of transport in another sample grown and measured at Purdue used in study of reentrant insulating phases by Deng et al. [77]. While the mobility of this sample was only 15 million, the 5/2 gap was $\Delta_{5/2}$=520mK. The 300mK $\nu$=5/2 resistivity, $\rho_{5/2}$, was 31.5$\Omega$/□, among the lowest we have ever measured. A comparison between the data of Xia et al. [69] in a 31 million sample and Deng et al. [77] in a 15 million sample is interesting. Note while the densities are slightly different (the Purdue sample is actually lower), both samples employ a doping well-design. Figure 14 shows data in the vicinity of $\nu$=5/2 in each sample. The factor of 2 discrepancy in mobility is not evident in the low temperature magnetotransport; in fact the $\nu$=5/2 gap is larger in the sample of Deng et al. [77]. In a more recent measurement using another doping-well sample with $\mu$=20x10$^6$cm$^2$/Vs and $\rho_{5/2}$ =35$\Omega$/□ , we have measured a $\nu$=5/2 excitation gap $\Delta_{5/2}$=570mK, the largest yet reported (Csathy and Manfra groups, to be published).

At this juncture, a few comments are necessary. At this time we do not claim that $\nu$=5/2 resistivity measured at T=0.3K is the best, or only, useful metric of 2DEG quality relevant to 2$^{nd}$ Landau level physics at very low temperatures; it is simply the one we chose to use in the absence of a clear alternative and the apparent lack of a strong correlation with zero-field mobility. In this review we have only shown a few examples of data in which low $\nu$=5/2 resistivity correlates with high quality transport in the 2$^{nd}$ Landau level at low temperatures. To make the argument convincing we need to show that samples with *higher* $\rho_{5/2}$ display *smaller* excitations gaps at $\nu$=5/2 and less well developed states at $\nu$=12/5. This study is currently underway but not yet complete. It is worth noting that we have grown several samples with designs that do not use the doping-well scheme but yet show high mobility (at or above 20 million). However these samples typically display higher $\rho_{5/2}$ than the doping-well design at T=0.3K. Measurement of their excitation gaps is part of our ongoing study. Perhaps a conservative, but fair, statement is that heterostructure design matters as much, if not more than, as zero-field mobility to the quality of transport features observed in the fractional quantum Hall regime. Simple comparisons of the mobility and excitation gaps as has been done extensively in the literature may overlook the material parameters most crucial observation of correlated states in a particular regime of filling factor. Clearly, efficient screening of the potential fluctuations due to remote silicon donor ions matters critically for the small-gapped states of the 2$^{nd}$ Landau level.

Other groups have also reported that the exact nature of the disorder must be considered when assessing 2DEG quality for the 2$^{nd}$ Landau level. Pan *et al*. [78] noted the gap at $\nu=5/2$ was less affected by short-range interface roughness scattering in field-effect transistor structures than by long-range potential fluctuations. Nuebler's assertion relating the 5/2 quasiparticle size and screening of remote silicon donors is especially relevant for the doping-well design [75]. As we have described, a significant amount of charge useful for screening may be available in the pure AlAs layers surrounding the doping wells [23]. Gamez *et al*. [79] have shown the significant over-doping of silicon doping can improve transport in around $\nu=5/2$ even in low mobility ($\sim 4 \times 10^6 cm^2/Vs$) non-doping-well samples. The Purdue group has recently begun an investigation of the impact of controllably adding different types of disorder (short range alloy scattering and long range Coulomb centers) into the doping-well design [80]. Using MBE, we can precisely control the amount and nature of the disorder introduced while keeping other important parameters such as 2DEG density and symmetry of the wavefunction fixed. As an example of the interesting results generated by this effort we note that we have taken the basic doping-well design, but instead of using a pure GaAs 30nm quantum well, we replaced it with an Al$_x$Ga$_{1-x}$As 30nm quantum well where x=0.0026. The 2DEG density was unaffected (it was $2.8 \times 10^{11} \text{cm}^{-2}$) but mobility using the $Al_xGa_{1-x}As$ well was only $2.7 \times 10^6 \text{cm}^2/\text{Vs}$. Yet $\nu=5/2$ remained fully quantized with an excitation gap of 200mK! Alloy disorder appears to impact mobility considerably while preserving a strong 5/2 state, again pointing to the importance of screening the remote impurities for 5/2 physics. The ability to controllably introduce disorder with MBE promises to yield important information about what types of disorder most relevant and how best to quantify that disorder.

**OPTIMIZATION OF SAMPLE DESIGN FOR SPECIFIC EXPERIMENTS**

Maximization of the $\nu=5/2$ excitation gap is just one of many possible parameters that may need to be optimized to tailor a sample to a specific experiment. The ability to controllably gate a sample, the 2DEG's proximity to the sample surface, and its temporal stability are also key factors for many experiments in the $2^{nd}$ Landau level. Moreover, there are many exciting experiments in quantum dots, bilayer quantum Hall systems, and two-dimensional hole systems that are not focused at all on the $\nu=5/2$ state and have completely different design criteria. We discuss some of these issues as they relate to MBE growth in the following section.

**Mesoscopic devices and quantum Hall interferometry**

So far we have only discussed measurements of the excitation gap at $\nu=5/2$ in 2DEG samples. Many of the most exciting proposals for exploring the non-Abelian properties of the $\nu=5/2$ state involve interrogation of edge states in confined geometries in which propagating edge states can be made to interfere [58-62]. Indeed, interference experiments are a principal method proposed to expose the non-Abelian statistics of the 5/2 quasiparticles. This requires device features near or below 1 micron. Devices of this type are typically realized using a combination of optical and e-beam lithography to define metallic top gates which in turn are used to electrostatically defined the geometry of the 2DEG. Many interesting experimental results have been reported [81-88] in this regime. The experiments of Willett *et al.* give the strongest indication of a non-Abelian state at ν=5/2 to date [89-93]. This class of experiments places some important constraints on heterostructure design as we now discuss.

Combining the practices commonly used to study mesoscopic physics with ultra-high quality 2DEGs designed to support strong fractional quantum Hall states in the 2$^{nd}$ Landau level is a challenging task. Historically, the best transport is usually observed in large area (~4mm by ~4mm) samples with minimal, if any, processing other than annealing of ohmic contacts. Clearly sample design is an important consideration for more complicated experimental geometries. The design must support robust fractional states with large excitation gaps at ν=5/2, 7/3, 8/3 and even possibly 12/5 while also demonstrating stable behavior with energized top gates. As we have detailed, the doping-well design yields the largest excitation gaps. Unfortunately, the doping-well design's distinguishing attribute, the presence of excess charge capable of screening the potential fluctuations of the remote silicon donors, also makes it problematic for top gated mesoscopic devices. Apparently, charge in the AlAs layers is not sufficiently mobile to make a significant contribution to near DC magnetotransport measurements, but it does impact behavior in structures with top gates. As reported by Rossler *et al*. [94] the doping-well design tends to produce hysteretic 2DEG density vs. gate voltage behavior. Perhaps more debilitating is that the 2DEG density is often observed to be temporally unstable; the 2DEG density changes as a function of time with a constant top gate voltage. Both behaviors of course make experiments extremely difficult if not impossible. Both behaviors also point to long time scale redistribution of charge within the heterostructure, mostly likely associated with excess, weakly mobile, charge between the gate and the principal 2DEG. While modifications to the doping-well design exist and have been implemented, to date, a fully satisfactory solution has not been found. The most stable configuration still places the dopant silicon atoms directly in the large band gap AlGaAs barrier. This method typically yields a slightly diminished, but still usable, excitation gap at ν=5/2 [89].

**Heterostructure design, GaAs quantum dots, and spin-qubits**

Coherent manipulation of localized spins in quantum dots is central to the realization of spin-based quantum computing [95]. Many experimental realizations of spin qubits are based on lithographically defined quantum dots built on top of a 2DEG in an AlGaAs/GaAs heterostructure. While extremely low disorder samples have not traditionally been required for quantum dot work (the electrons are localized on a length scale much shorter than a typical mean free path), issues related to disorder, gating stability, and electronic noise generated within the AlGaAs/GaAs heterostructure are becoming increasingly important as experimentalists attempt to scale-up to coherent control of multiple spin qubits [95-105].

Heterostructures designed for spin qubits tend to differ markedly from those used to explore the fractional quantum Hall effect in the 2$^{nd}$ Landau level. For spin qubits, the 2DEG is placed close to surface, typically between 50 and 100nm, to aid electrostatic confinement with surface gates. Working in this regime will inherently limit 2DEG mobility as a significant number of ionized donor impurities will be placed in close proximity to the conducting channel to account for band bending and surface Fermi-level pinning. Nevertheless 2DEG mobility exceeding $4x10^6 cm^2$/Vs can still be achieved in these shallow 2DEG designs. As time-dependent potential fluctuations can cause decoherence, a primary heterostructure design objective is material the will produce "quiet" devices. While is generally believed that time-dependent fluctuations are associated with charge tunneling into and out of donor impurities, a complete understanding of noise and how to mitigate it in AlGaAs/GaAs heterostructures has yet to be achieved [106, 107]. In our laboratory, heterostructures designed for operation as spin qubits have focused on two approaches. We grow modulation-doped single heterojunctions with high aluminum content barriers (x>0.35) and uniformly distributed silicon dopants in order to understand the origins of charge noise. Interestingly this structure design, one used from the earliest days in the development of AlGaAs/GaAs 2DEGs, seems to produce the lowest levels of charge noise among current designs that utilize silicon doping to produce a 2DEG. Another approach currently under development involves the exclusion of silicon dopants altogether, relying instead on the use of an insulated gate to generate carriers at an AlGaAs/GaAs interface through a strong electric field effect. This device design shares many similarities with approaches commonly found in mainstream silicon technology and offers the possibility of reduced charge noise if indeed the silicon dopant atoms are the principal source of noise.

**Bilayer two-dimensional electron systems**

One of the most fruitful directions in current research centers on what happens when two high quality 2DEGs are brought into close enough proximity to form a bilayer electronic system. At high magnetic fields, many new states not found in single-layer samples have been discovered [108-112] associated with the extra degree of freedom, the bilayer layer-index, and strong electron-electron interactions in low disorder samples. In this system the most spectacular behavior is condensation of excitons formed from electrons in one layer with holes in the other layer at total filling factor $\nu_T=1$ and small values of d/l, where d is the center-to-center distance between the two GaAs quantum wells and l is the magnetic length, $l = \sqrt{\hbar/eB}$. This state is analogous to a BCS-like pairing of Cooper pairs, supporting unusual superfluid behavior [113-117]. Bilayer electron samples place significant demands on the MBE grower. A typical structure consists of 18nm GaAs quantum wells separated by a 10nm $Al_{0.9}Ga_{0.1}As$ barrier. Total electron density is usually around ~$1x10^{11}cm^{-2}$, shared equally between the two layers. The disorder introduced by the thin, high aluminum content, barrier limits mobility to about $1x10^6 cm^2/Vs$. As grown, the bilayer layers often must be gated down to lower density to explore the $v_T=1$ state. This process further reduces bilayer quality. Recently, lower density samples with improved quality have been achieved [118, 119]. Further improvements in sample quality at densities significantly below $1x10^{11}cm^{-2}$ are expected to yield new results at fractional values of total filling. Several strategies to improve low density and low disorder bilayer heterostructures are currently under development in our laboratory.

**Two-dimensional hole systems**

A two-dimensional system of valence band holes can also be formed at the AlGaAs/GaAs interface. The two-dimensional hole system (2DHS) has several properties that distinguish it from the more thoroughly studied 2DEG. Principal among these are larger and tunable effective mass, stronger spin-orbit coupling, and absence of direct hyperfine coupling to the nuclear field of the GaAs host lattice. The MBE grower has substantial flexibility as several properties including effective mass and spin-orbit coupling can be altered by heterostructure design, thus affording an opportunity to examine the impact of these material parameters on correlated-state formation. A principal drawback that has limited use of 2DHSs in this area has been the generally lower quality (i.e. mobility) exhibited by 2DHSs when compared to 2DEGs. Traditional p-type dopants for GaAs include beryllium which diffuses significantly in the GaAs lattice at MBE growth temperatures, limiting 2DHS quality [14]. Historically, the highest mobility 2DHS were grown on the (311)A face of GaAs where silicon can be incorporated as an acceptor [120-122]. The use of efficient carbon doping techniques on the (100) face of GaAs has resulted in extremely high mobility and isotropic 2DHSs [123, 124]. The highest mobility is now over $2.5 \times 10^6$ cm$^2$/Vs. These samples have been used to study the impact of spin-orbit coupling on quantum Hall nematic phases [125], the 2D metal-to-insulator transition [126], fractional quantum Hall states in the 2$^{nd}$ Landau level [127, 130], and the unusual limits to mobility in 2DHSs [128, 129]. It is speculated that the lack of direct hyperfine coupling to the GaAs nuclei could be useful for spin-based quantum computing.

**OUTLOOK**

The two-dimensional electron gas in AlGaAs/GaAs heterostructures has been studied for over thirty years, yet the subject remains vibrant. Non-Abelian fractional quantum Hall states, localized spins in quantum dots, and BCS-like condensation of excitons in electron bilayers are just a few prominent examples of physics presently explored with GaAs 2DEGs grown by MBE. Interestingly, some of these phenomena discovered in GaAs 2DEGs are now leading candidates for distinct implementations of quantum computing. Furthermore, study of the GaAs 2DEG forms the basis for our developing understanding of newly appreciated topological phases occurring in several new, but less well-developed, low-dimensional material systems. We expect that further improvement of GaAs 2DEG quality through innovations in MBE growth will lead to the discovery of new physics and attendant technological innovations. Based on existing experimental data, a path towards higher quality heterostructures based on improvement of MBE starting material quality has been outlined. Methodologies developed to produce higher quality GaAs/AlGaAs 2DEGs will also undoubtedly provide a template for materials purification and heterostructure design breakthroughs in other frontier solid-state systems.

**Acknowledgements**

I thank Loren Pfeiffer and Ken West; they taught me much of what I know about MBE. Bob Willett, Rafi de-Picciotto, Kirk Baldwin and Steve Simon provided a fertile intellectual environment for collaboration at Bell Laboratories. Bob and Rafi in particular taught me how to perform careful transport measurements. I have learned much about scattering in 2DEGs from Sankar Das Sarma. I thank my first batch of graduate students at Purdue: John Watson, Geoff Gardner and Sumit Mondal. I would also like to acknowledge Gabor Csathy. All of the new Purdue ultra-low temperature transport data in the $2^{nd}$ Landau level discussed in this review is the product of the Csathy group.

**REFERENCES**

1) D. C. Tsui, H. L. Stormer, and A. C. Gossard, Phys. Rev. Lett. **48**, 1559 (1982)

2) D. H. Torchinsky, F. Mahmood, A. T. Bollinger, I. Bozovic, and N. Gedik, Nature Materials **12**, 387 (2013)

3) S. Oh, T. A. Crane, D. J. Van Harlingen, and J. N. Eckstein, Phys. Rev. Lett. **96**, 107003 (2006)

4) A. Ohtomo and H. Y. Hwang, Nature **427**, 423 (2004)

5) C. M. Brooks, L. Fitting Kourkoutis, T. Heeg, J. Schubert, D. A. Muller, and D. G. Schlom, Appl. Phys. Lett. **94**, 162905 (2009)

6) G. Wang, X. Zhu, Y. Sun, Y. Li, T. Zhang, J. Wen, X. Chen, K. He, L. Wang, X. Ma, J. Jia, S. Zhang, and Q. Xue, Adv. Mater. **23**, 2929 (2011)

7) M. Konig, S. Wiedmann, C. Brune, A. Roth, H. Buhmann, L. W. Molenkamp, X. Qi, and S. Zhang, Science **318**, 767 (2007)

8) A. Y. Cho and J. R. Author, Prog. Solid State Chem. **10**, 157 (1975)

9) J. Y. Tsao, *Materials Fundamentals of Molecular Beam Epitaxy* (Academic Press Inc., San Diego, 1993)

10) A. Pimpinelli and J. Villain, *Physics of Crystal Growth* (Cambridge University Press, Cambridge, 1998)

11) M. A. Herman and H. Sitter, *Molecular Beam Epitaxy: Fundamentals and Current Status* (Springer-Verlag, Berlin, 1996 2$^{nd}$ Edition)

12) N. N. Ledentsov, *Growth Processes and Surface Phase Equilibria in Molecular Beam Epitaxy* (Springer-Verlag, Berlin, 1999)

13) R. F. C. Farrow ed., *Molecular Beam Epitaxy, Applications to Key Materials* (Noyes Publications, Park Ridge, 1995)

14) E. F. Schubert, *Doping in III-V Semiconductors* (Cambridge University Press, Cambridge, 1995)

15) M. R. Melloch, Thin Solid Films, **21**, 74 (1993)

16) L. N. Pfeiffer, K. W. West, R. L. Willett, H. Akiyama, and L. P. Rokhinson, Bell Labs Tech. Journal, **10**, 151 (2005)

17) H. T. Chou, S. Luscher, D. Goldhaber-Gordon, M. J. Manfra, A. M. Sergent, K. W. West and R. J. Molnar, Appl. Phys. Lett. **86**, 073108 (2005)

18) M. J. Manfra, K. W. Baldwin, A. M. Sergent, K. W. West, R. J. Molnar and J. Caissie, Appl. Phys. Lett. **85**, 5394 (2004)

19) M. J. Manfra, K. W. Baldwin, A. M. Sergent, R. J. Molnar and J. Caissie, Appl. Phys. Lett. **85**, 1722 (2004)

20) R. Dingle, H. L. Stormer, A. C. Gossard and W. Wiegmann, Appl. Phys. Lett. **33**, 665 (1978)

21) H. L. Stormer, R. Dingle, A. C. Gossard, W. Wiegmann and M. D. Sturge, Solid State Comm. **29**, 705 (1979)

22) J. P. Eisenstein, K. B. Cooper, L. N. Pfeiffer, and K. W. West, Phys. Rev. Lett. **88**, 076801 (2002)

23) V. Umansky, M. Heiblum, Y. Levinson, J. Smet, J. Nubler and M. Dolev, J. Cryst. Growth **311**, 1658 (2009)

24) E. H. Hwang and S. Das Sarma, Phys. Rev. B **77**, 235437 (2008)

25) http://www.veeco.com/promos/mbe/manfra/2011_aug_manfra.aspx

26) L. N. Pfeiffer, K. W. West, H. L. Stormer, and K. W. Baldwin, Appl. Phys. Lett. **55**, 1888 (1989)

27) The Purdue system is equipped with the IE 514 "extractor" gauge and associated electronics from Oerlikon Lybold Vacuum Inc.

28) The Purdue system is equipped with two RGAs; a 200 amu unit from Stanford Research Systems Inc. and another 200 amu unit from Ametek Process Instruments. Both systems specify a minimum detectable partial pressure of $5 \times 10^{-14}$ torr.

29) Y. Okada, T. Sugaya, S. Ohta, T. Fujita, and M. Kawabe, Jpn. J. Appl. Phys. **34**, 238 (1995)

30) Y. Okada, T. Fujita, and M. Kawabe, Appl. Phys. Lett. **67**, 676 (1995)

31) Z. Y. Zhou, C. X. Zheng, W. X. Tang, D. E. Jesson, and J. Tersoff, Appl. Phys. Lett. **97**, 121912 (2010)

32) V. Umansky, R. de-Picciotto, and M. Heiblum, Appl. Phys. Lett. **71**, 683 (1997)

33) Depending on the specific element, secondary ion mass spectrometry usually has a noise limit no lower than $10^{-15}/cm^3$.

34) D. V. Lang, J. Appl. Phys. **45**, 3023 (1974)

35) P. Blood and J. J. Harris, J. Appl. Phys. **56**, 993 (1984)

36) E. C. Larkins, E. S. Hellman, D. G. Schlom, J. S. Harris Jr., M. H. Kim and G. E. Stillman, Appl. Phys. Lett. **49** 391 (1986)

37) N. Chand, R. C. Miller, A. M. Sergent, S. K. Sputz, and D. V. Lang, Appl. Phys. Lett. **52** 1721 (1988)

38) N. Chand, T. D. Harris, S. N. G. Chu, E. E. Becker, A. M. Sergent, M. Schnoes and D. V. Lang, J. Crystal Growth **111**, 20 (1991)
38

39) C. R. Stanley, M. C. Holland, A. H. Kean, J. M. Chamberlain, R. T. Grimes and M. B. Stanaway, J. Crystal Growth **111**, 14 (1991)

40) J. F. O'Hanlon, *A User's Guide to Vacuum Technology – 2$^{nd}$ Edition* (Wiley, New York 1989)

41) S. Schmult, S. Taylor, and W. Dietsche, J. Crystal Growth **311**, 1655 (2009)

42) D. J. Chadi and K. J. Chang, Phys. Rev. Lett. **61**, 873 (1988); P. M. Mooney, J. Appl. Phys. **67**, R1 (1990)

43) D. J. Chadi and K. J. Chang, Phys. Rev. B **39**, 10063 (1989)

44) N. Chand, *et al*. Phys. Rev. B **30**, 4481 (1984)

45) K. J. Friedland, R. Hey, H. Kostial, R. Klann, and K. Ploog, Phys. Rev. Lett. **77**, 4616 (1996)

46) R. B. Laughlin, Phys. Rev. Lett. **50**, 1395 (1983)

47) F. D. M. Haldane, Phys. Rev. Lett. **51**, 605 (1983)

48) B. I. Halperin, Phys. Rev. Lett. **52**, 1583 (1984)

49) S. Das Sarma, and A. Pinzcuk (eds.) *Perspectives in quantum Hall effects: novel quantum liquids in low-dimensional semiconductor structures* (Wiley, New York 1997)

50) R. Prange and S. M. Girvin (eds.) *The Quantum Hall effect* (Springer-Verlag, New York 1990)

51) G. Moore and N. Read, Nucl. Phys. B **360**, 362 (1991)

52) M. Levin, B. I. Halperin, and B. Rosenow, Phys. Rev. Lett. **99**, 236806 (2007); S. S. Lee, S. Ryu, C. Nayak, and M. P. A. Fisher, Phys. Rev. Lett. 99, 236807 (2007)

53) J. K. Jain, Phys. Rev. Lett. **63**, 199 (1989)

54) M. Greiter, X. G. Wen and F. Wilczek, Nucl. Phys. B **374**, 567 (1992)

55) N. Read, and D. Green, Phys. Rev. B **61**, 10267 (2000)

56) B. I. Halperin, P. A. Lee and N. Read, Phys. Rev. B **47**, 7312 (1993)

57) N. Read and E. Rezayi, Phys. Rev. B **54**, 8084 (1996)

58) S. Das Sarma, M. Freedman, and C. Nayak, Phys. Rev. Lett. **94**, 16802 (2005)

59) C. Nayak, S. H. Simon, A. Stern, M. Freedman, and S. Das Sarma, Rev. Mod. Phys. **80**, 1083 (2008)

60) P. Bonderson, A. Kitaev, and K. Shtengel, Phys. Rev. Lett. **96**, 016803 (2006)

61) A. Stern and B. I. Halperin, Phys. Rev. Lett. **96**, 016802 (2006)

62) A. Stern, F. von Oppen, and E. Mariani, Phys. Rev. B **70**, 205338 (2004)

63) R. L. Willett, J. P. Eisenstein, H. L. Stormer, D. C. Tsui, A. C. Gossard, and J. H. English, Phys. Rev. Lett. **59**, 1776 (1987)

64) J. P. Eisenstein, R. L. Willett, H. L. Stormer, L. N. Pfeiffer and K. W. West, Surf. Science **229**, 31 (1990)

65) W. Pan, J. S. Xia, V. Shvarts, D. E. Adams, H. L. Stormer, D. C. Tsui, L. N. Pfeiffer, K. W. Baldwin and K. W. West, Phys. Rev. Lett. **83**, 3530 (1999)

66) M. Storni, R. H. Morf, and S. Das Sarma, Phys. Rev. Lett. **104**, 076803 (2010)

67) R. H. Morf and N. d'Ambrumenil, Phys. Rev. B **68**, 113309 (2003)

68) R. H. Morf, N. d'Ambrumenil and S. Das Sarma, Phys. Rev. B **66**, 075408 (2002)

69) J. S. Xia, W. Pan, C. L. Vicente, E. D. Adams, N. S. Sullivan, H. L. Stormer, D. C. Tsui, L. N. Pfeiffer, K. W. Baldwin, and K. W. West, Phys. Rev. Lett. **93**, 176809 (2004)

70) N. Read, and E. Rezayi, Phys. Rev. B **59**, 8084 (1999)

71) W. Pan, J. S. Xia, H. L. Stormer, D. C. Tsui, C. L. Vincente, E. D. Adams, N. S. Sullivan, L. N. Pfeiffer, K. W. Baldwin, and K. W. West, Phys. Rev. B **77**, 075307 (2008)

72) A. Kumar, G. A. Csathy, M. J. Manfra, L. N. Pfeiffer and K. W. West, Phys. Rev. Lett. **105**, 246808 (2010); N. Deng, A. Kumar, M. J. Manfra, L. N. Pfeiffer, K. W. West, and G. A. Csathy, Phys. Rev. Lett. **108**, 086803 (2012)

73) H. C. Choi, W. Kang, S. Das Sarma, L. N. Pfeiffer and K. W. West, Phys. Rev. B **77**, 081301 (2008)

74) C. R. Dean, B. A. Piot, P. Hayden, S. Das Sarma, G. Gervais, L. N. Pfeiffer, and K. W. West, Phys. Rev. Lett. **100**, 146803 (2008)

75) J. Nuebler, V. Umansky, R. Morf, M. Heiblum, K. von Klitzing, and J. Smet, Phys. Rev. B **81**, 035316 (2010)

76) N. Samkharadze, J. D. Watson, G. Gardner, M. J. Manfra, L. N. Pfeiffer, and K. W. West, and G. A. Csathy, Phys. Rev. B **84**, 121305(R) (2011)

77) N. Deng, J. D. Watson, L. P. Rokhinson, M. J. Manfra, and G. A. Csathy, Phys. Rev. B **86**, 201301(R) (2012)

78) W. Pan, N. Masuhara, N. S. Sullivan, K. W. Baldwin, K. W. West, L. N. Pfeiffer and D. C. Tsui, Phys. Rev. Lett. **106**, 206806 (2011)

79) G. Gamez, and K. Muraki, arXiv: 1101.5856v1

80) G. Gardner, J. D. Watson, S. Mondal, N. Deng, G. A. Csathy and M. J. Manfra, Appl. Phys. Lett. **102**, 252103 (2013)

81) A. Kou, C. M. Marcus, L. N. Pfeiffer, and K. W. West, Phys. Rev. Lett. **108**, 256803 (2012)

82) D. T. McClure, W. Chang, C. M. Marcus, L. N. Pfeiffer, and K. W. West, Phys. Rev. Lett. **108**, 256804 (2012)

83) D. T. McClure, Y. Zhang, B. Rosenow, E. M. Levenson-Falk, C. M. Marcus, L. N. Pfeiffer, and K. W. West, Phys. Rev. Lett. **103**, 206806 (2009)

84) Y. Zhang, D. T. McClure, E. M. Levenson-Falk, C. M. Marcus, L. N. Pfeiffer, and K. W. West, Phys. Rev. B **79**, 241304(R) (2009)

85) I. P. Radu, J. B. Miller, C. M. Marcus, M. A. Kastner, L. N. Pfeiffer, and K. W. West Science **320**, 899 (2008)

86) J. B. Miller, I. P. Radu, D. M. Zumbuhl, E. M. Levenson-Falk, M. A. Kaster, C. M. Marcus, L. N. Pfeiffer, and K. W. West, Nat. Phys. **3**, 561 (2007)

87) M. Dolev, M. Heiblum, V. Umansky, A. Stern, and D. Mahalu, Nature **452**, 829 (2008)

88) N. Ofek, A. Bid, M. Heiblum, A. Stern, V. Umansky, and D. Mahalu, Proc. Nat. Acad. Sci. **107**, 5276 (2010)

89) R. L. Willett, L. N. Pfeiffer, and K. W. West, Phys. Rev. B **82**, 205301 (2010)

90) R. L. Willett, L. N. Pfeiffer, and K. W. West, Proc. Nat. Acad. Sci. **106**, 8853 (2009)

91) R. L. Willett, M. J. Manfra, L. N. Pfeiffer, and K. W. West, Appl. Phys. Lett. **91**, 052105 (2007)

92) R. L. Willett, C. Nayak, K. Shtengel, L. N. Pfeiffer, and K. W. West, arXiv:1301.2639v1

93) R. L. Willett, L. N. Pfeiffer, K. W. West, and M. J. Manfra, arXiv:1301.2594v1

94) C. Rossler, T. Feil, R. Mensch, T. Ihn, K. Ensslin, D. Schuh, and W. Wegscheider, New J. Phys. **12**, 043007 (2010)

95) C. Kloeffel, and D. Loss, Annu. Rev. Condens. Matter Phys. **4**, 51 (2013)

96) O. E. Dial, M. D. Shulman, S. P. Harvey, H. Blum, V. Umansky, and A. Yacoby, Phys. Rev. Lett. **110**, 146804 (2013)

97) M. D. Shulman, O. E. Dial, S. P. Harvey, H. Bluhm, V. Umansky, and A. Yacoby, Science **336**, 202 (2012)

98) L. Trifunovic, O. Dial, M. Trif, J. R. Wootton, R. Abebe, A. Yacoby, and D. Loss, Phys. Rev. X **2**, 011006 (2012)

99) H. Bluhm, S. Foletti, I. Neder, M. Rudner, D. Mahalu, V. Umansky, and A. Yacoby, Nat. Phys. **7**, 109 (2011)

100) H. Bluhm, S. Foletti, D. Mahalu, V. Umansky, and A. Yacoby, Phys. Rev. Lett. **105**, 216803 (2010)

101) J. Medford, L. Cywinski, C. Barthel, C. M. Marcus, M. P. Hanson, and A. C. Gossard, Phys. Rev. Lett. **108**, 086802 (2012)

102) I. van Weperen, B. D. Armstrong, E. A. Laird, J. Medford, C. M. Marcus, M. P. Hanson, and A. C. Gossard, Phys. Rev. Lett. **107**, 030506 (2011)

103) C. Barthel, J. Medford, C. M. Marcus, M. P. Hanson, and A. C. Gossard, Phys. Rev. Lett. **105**, 266808 (2010)

104) F. R. Braakman, P. Barthelemy, C. Reichl, W. Wegscheider, and L. M. K. Vandersypen, Nature Nanotech. **8**, 432 (2013)

105) V. Srinivasa, K. C. Nowack, M. Shafiei, L. M. K. Vandersypen, and J. M. Taylor, Phys. Rev. Lett. **110**, 196803 (2013)

106) C. Buizert, F. H. L. Koppens, M. Pioro-Ladriere, H-P. Tranitz, I. T. Vink, S. Tarucha, W. Wegscheider, and L. K. Vandersypen, Phys. Rev. Lett. 101, 226603 (2008)

107) K. Hitachi, T. Ota, and K. Muraki, arXiv:1305.6701v1

108) Y. W. Suen, L. Engel, M. B. Santos, M. Shayegan, and D. C. Tsui, Phys. Rev. Lett. **68**, 1379 (1992)

109) J. P. Eisenstein, G. S. Boebinger, L. N. Pfeiffer, K. W. West, and S. He, Phys. Rev. Lett. **68**, 1383 (1992)

110) I. B. Spielman, J. P. Eisenstein, L. N. Pfeiffer, and K. W. West, Phys. Rev. Lett. **84**, 5808 (2000)

111) M. Kellogg, I. B. Spielman, J. P. Eisenstein, L. N. Pfeiffer, and K. W. West, Phys. Rev. Lett. **88**, 126804 (2002)

112) M. Kellogg, J. P. Eisenstein, L. N. Pfeiffer, and K. W. West, Phys. Rev. Lett. **93**, 036801 (2004)

113) I. B. Spielman, J. P. Eisenstein, L. N. Pfeiffer, and K. W. West Phys. Rev. Lett. **87**, 036803 (2001)

114) A. R. Champange, A. D. K. Finck, J. P. Eisenstein, L. N. Pfeiffer, and K. W. West, Phys. Rev. B **78**, 205310 (2008)

115) S. Misra, N. C. Bishop, E. Tutuc, and M. Shayegan, Phys. Rev. B 77, 161301(R) (2008)

116) Y. Yoon, L. Tiemann, S. Schmult, W. Dietsche, K. von Klitzing, and W. Wegscheider, Phys. Rev. Lett. **104**, 116802 (2010)

117) D. Nandi, A. D. K. Finck, J. P. Eisenstein, L. N. Pfeiffer, and K. W. West, Nature **488**, 481 (2012)

118) S. Schmult, L. Tiemann, W. Dietsche, and K. von Klitzing, J. Vac. Sci. Technol. B **28**, C3C1 (2010)

119) Similar quality low-density bilayer samples have been grown at Purdue and are currently under test.

120) M. B. Santos, Y. W. Suen, M. Shayegan, Y. P. Li, L. W. Engel, D. C. Tsui, Phys. Rev. Lett. **8**, 1188 (1992)

121) S. J. Papadakis, E. P. De Poortere, H. C. Manoharan, M. Shayegan, and R. Winkler, Science **283**, 2056 (1999)

122) S. J. Papadakis, E. P. De Poortere, M. Shayegan, and R. Winkler, Phys. Rev. Lett. **84**, 5592 (2000)

123) M. J. Manfra, L. N. Pfeiffer, K. W. West, R. de Picciotto, and K. W. Baldwin, Appl. Phys. Lett. **86**, 162106 (2005)

124) G. Gerl, S. Schmult, H.-P. Tranitz, C. Mitzkus, and W. Wegscheider, Appl. Phys. Lett. **86**, 252105 (2005)

125) M. J. Manfra, R. de Picciotto, Z. Jiang, S. H. Simon, L. N. Pfeiffer, K. W. West, and A. M. Sergent, Phys. Rev. Lett. **98**, 206804 (2007)

126) M. J. Manfra, E. H. Hwang, S. Das Sarma, L. N. Pfeiffer, K. W. West, and A. M. Sergent, Phys. Rev. Lett. **99**, 236402 (2007)

127) A. Kumar, N. Samkharadze, G. A. Csathy, M. J. Manfra, L N. Pfeiffer, and K. W. West, Phys. Rev. B **83**, 201305(R) (2011)

128) J. D. Watson, S. Mondal, G. Gardner, G. A. Csathy, and M. J. Manfra, Phys. Rev. B **85**, 165301 (2012)

129) J. D. Watson, S. Mondal, G. A. Csathy, M. J. Manfra, E. H. Hwang, S. Das Sarma, L. N. Pfeiffer, and K. W. West, Phys. Rev. B **83**, 241305(R) (2011)

130) S. P. Koduvayur, Y. Lyanda-Geller, S. Khlebnikov, G. A. Csathy, M. J. Manfra, L. N. Pfeiffer, K. W. West, and L. P. Rokhinson, Phys. Rev. Lett. **106**, 016804 (2011)

131)     Nextnano3 simulator, ©1998-2008, Walter Schottky Institute

[http://www.nextnano.de/nextnano3/index.htm]

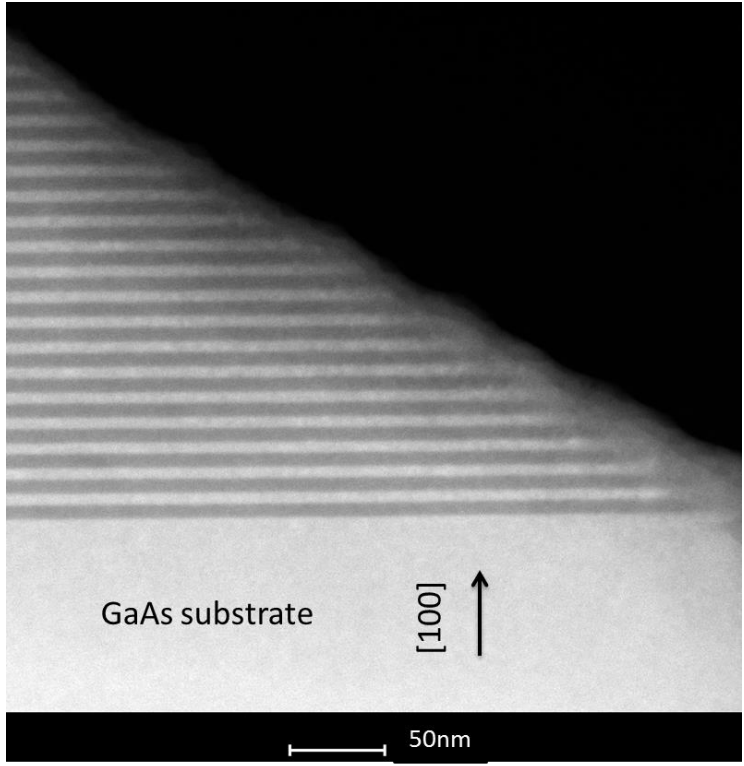

Figure 1: Cross-sectional TEM image of a 7nm AlAs, 5nm GaAs 50 period superlattice grown at Purdue. The bright regions are the GaAs layers while the darker regions correspond to AlAs.

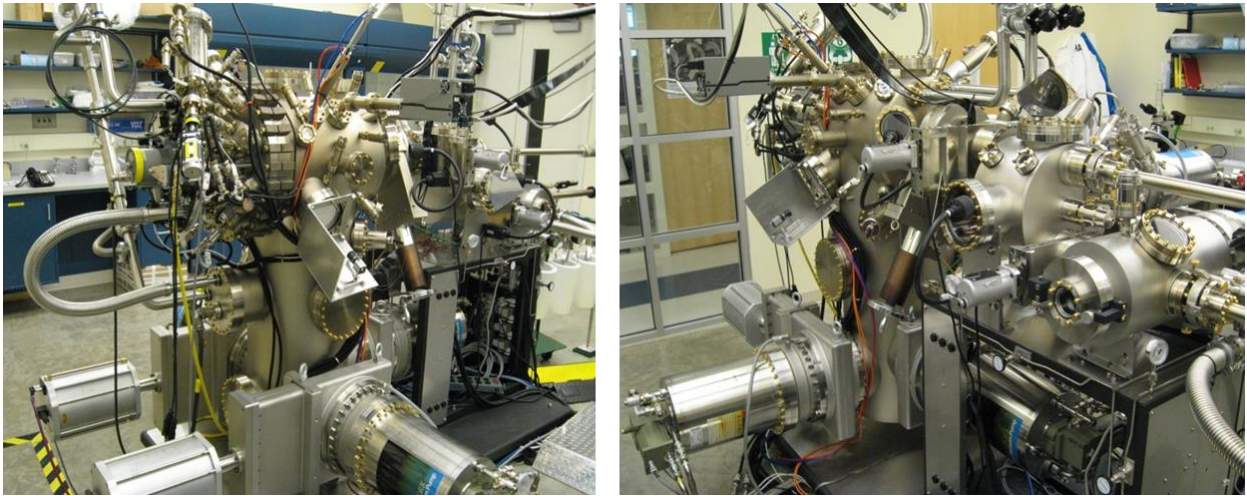

Figure 2: GaAs MBE system operating at Purdue. The image on the left shows the main growth chamber including source flange, in-situ diagnostic tools, and pumping configuration. The load-lock sample entry chamber and the intermediate sample outgassing chamber are visible in the image on the right. Each of these ancillary chambers is pumped by its own dedicated closed-cycle helium cryopump.

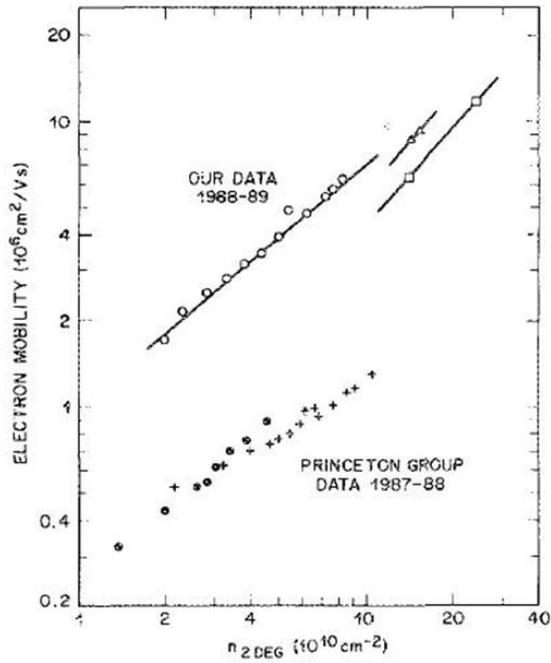 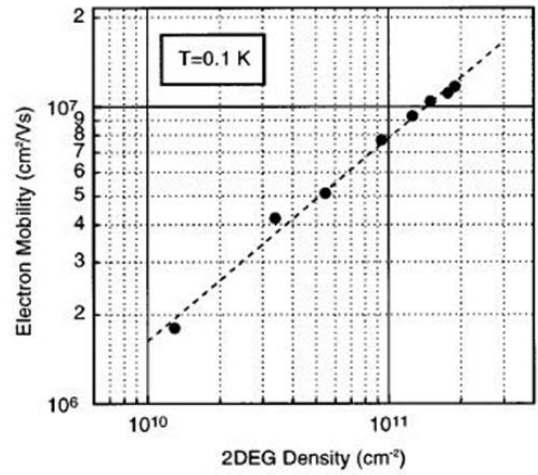

Figure 3: Dependence of low-temperature mobility vs. 2D density in modulation-doped single interfaces with large setbacks. Mobility scales as $\sim n_e^\alpha$ with $\alpha \sim 0.7$. (Data on the left after Ref. [26] and data on the right after Ref. [32])

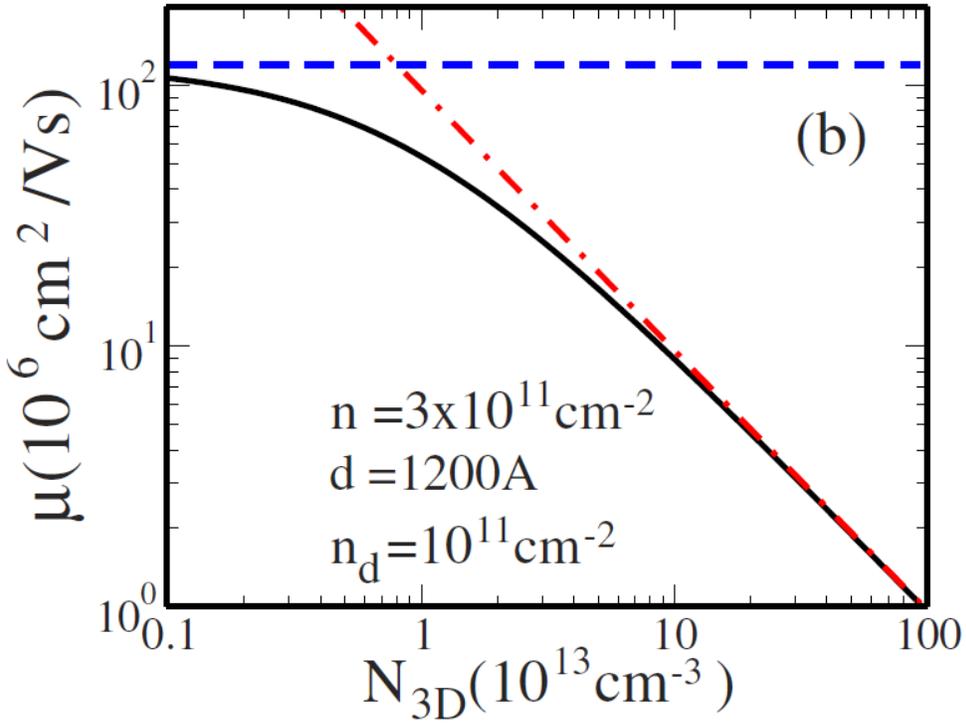

Figure 4: Calculated mobility as a function of background impurity density for 2D electron density n=3x10$^{11}$cm$^{-2}$, donor density n$_d$=10$^{11}$cm$^{-2}$ and setback d=120nm. The dashed (dot-dashed) curves show remote (background) charged impurity contribution. (After Hwang and Das Sarma [24])

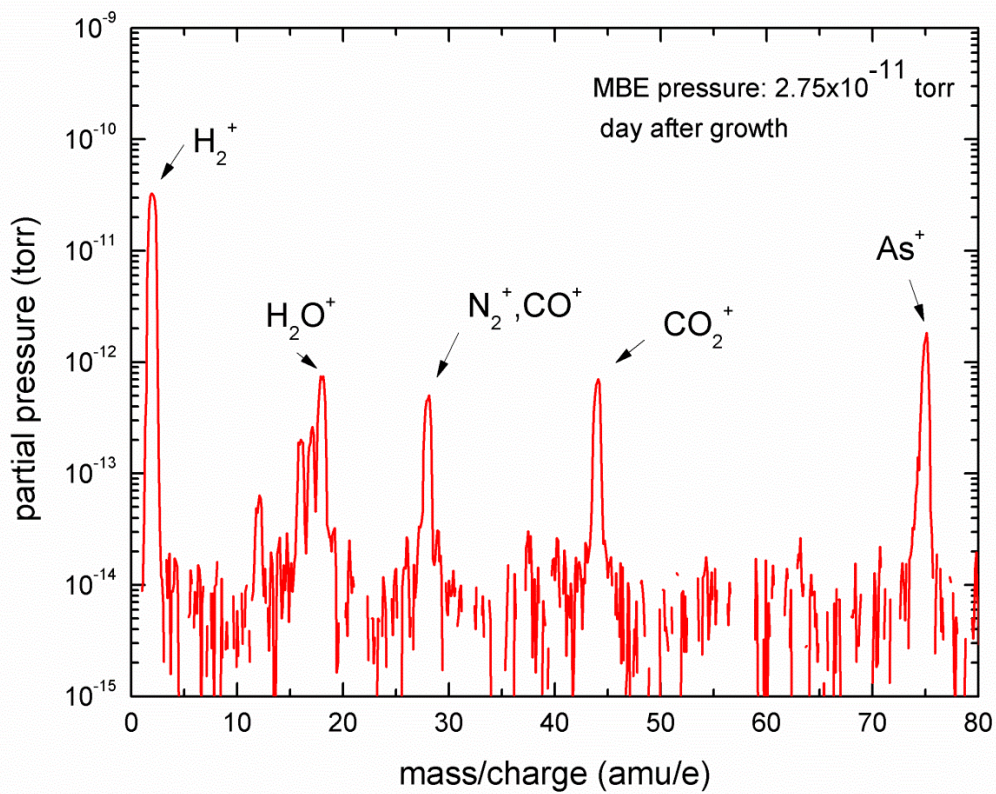

Figure 5: Spectrum from a residual gas analyzer attached to the growth chamber of the Purdue MBE system. The spectrum was taken a day after the MBE was used for growth and looks virtually identical to the spectrum after initial system bake-out, except for the presence of arsenic at amu/e=75.

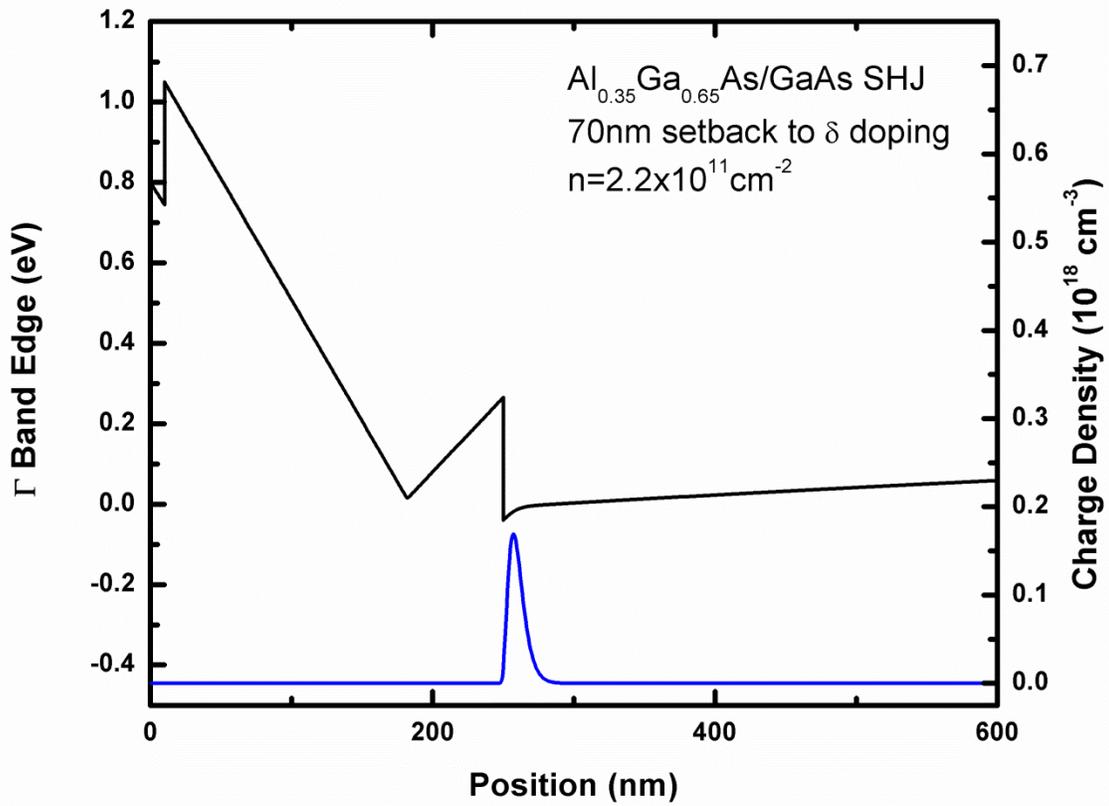

Figure 6: Position dependence of conduction band edge and charge density profile in a simple SHJ 2DEG. Calculated with NextNano3 [131].

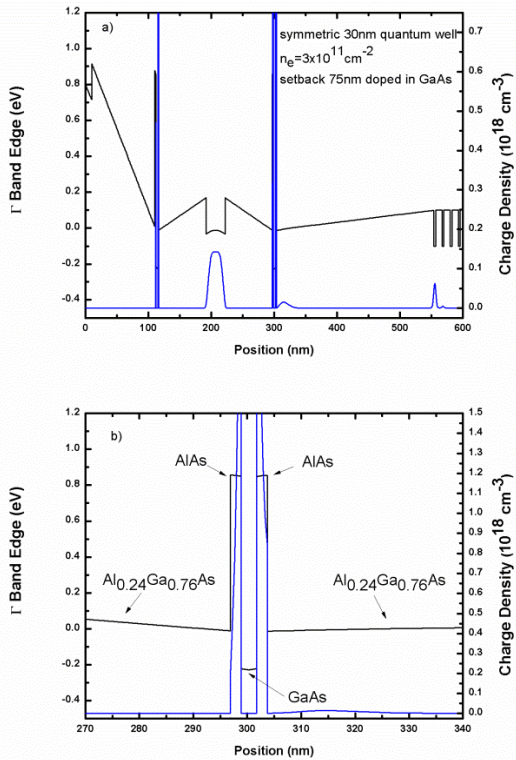

Figure 7: a) conduction band edge and charge density profile for a modern high mobility quantum well. Rather than delta-doping with silicon directly in the barrier, the silicon impurities are located in the center of narrow 3nm GaAs well surrounded by 2nm of AlAs. Charge is transferred not only to the primary 2DEG in the 30nm GaAs quantum well, but also to the X point band edge of the AlAs barriers as indicated in b). b) $\Gamma$ point conduction band edge and free charge density in the immediate vicinity of the "doping-well" located 75nm below the edge of the primary 30nm GaAs quantum well. Calculated with NextNano3 [131].

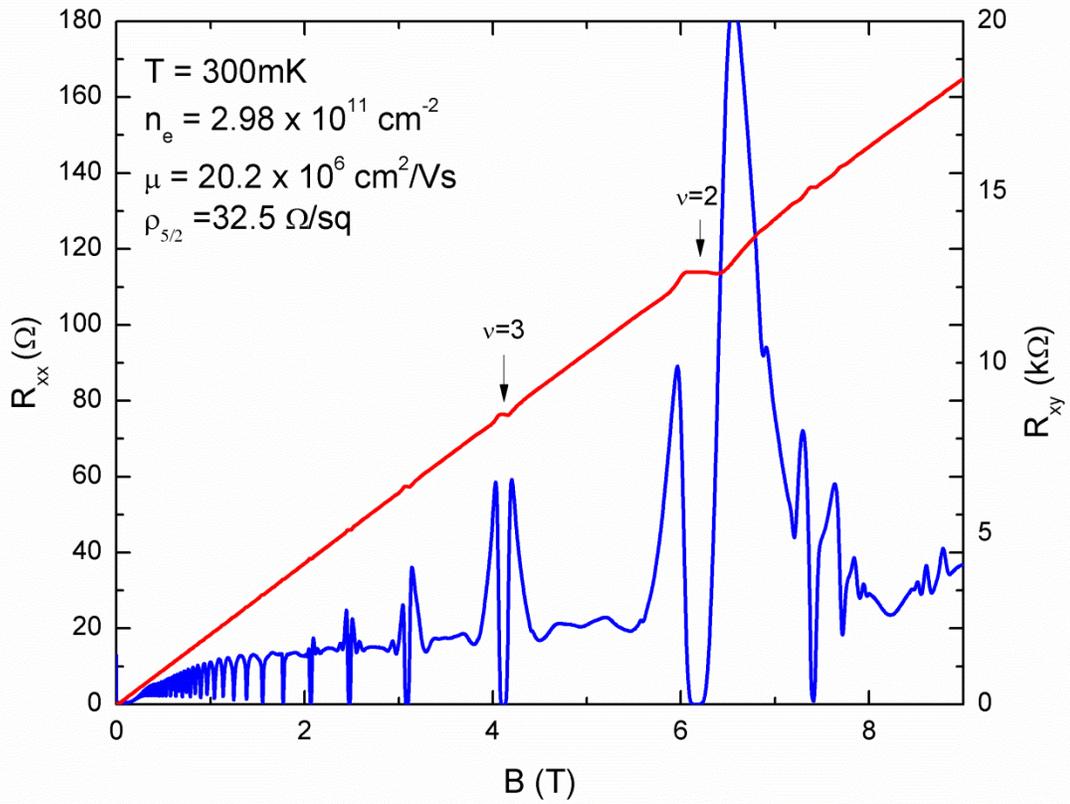

Figure 8: Longitudinal and Hall resistance in a doping-well sample at T=0.3K. The absence of significant parallel conduction from the doping-well region is evidenced by the strong zeroes in $R_{xx}$ in the integer quantum Hall states.

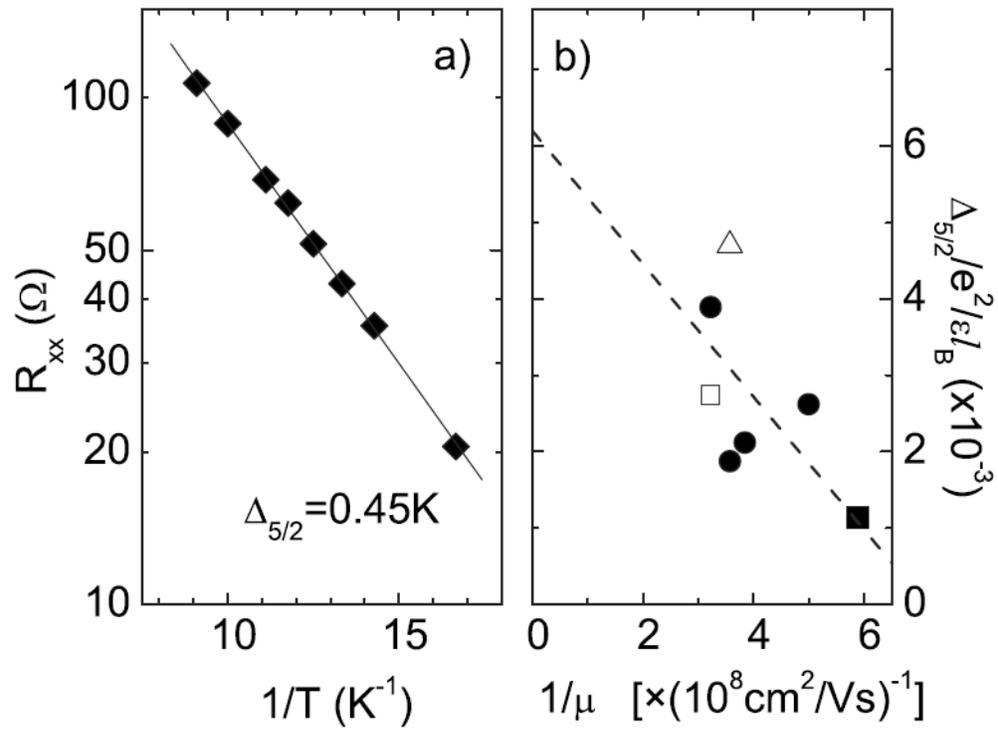

Figure 9: (a) Arrhenius plot for the $R_{xx}$ minimum at $\nu=5/2$. The line is a linear fit. (b) Normalized energy gap for five samples of different mobility. The line shows a linear fit to the data points. (Plot taken from Ref. [71])

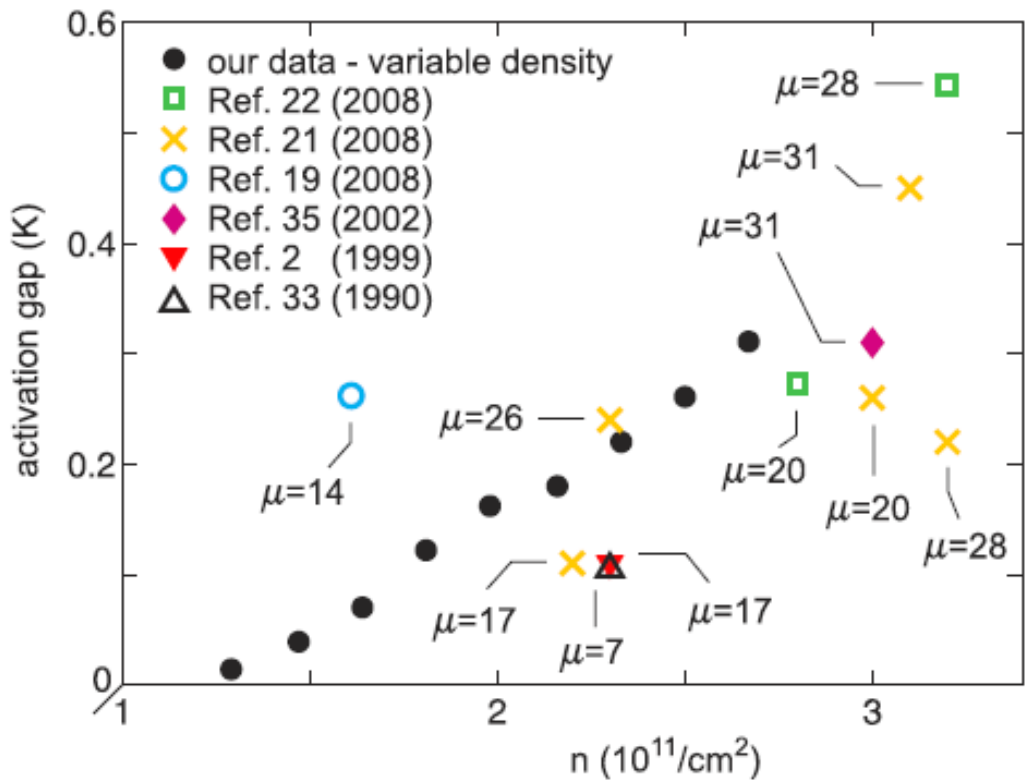

Figure 10: Figure from Nuebler *et al*., Ref. [75]. Comparison of 5/2 gaps measured in Ref. [75] to literature values. Mobility ($\mu$) of the samples used in the literature are given in units of $10^6 cm^2/Vs$.

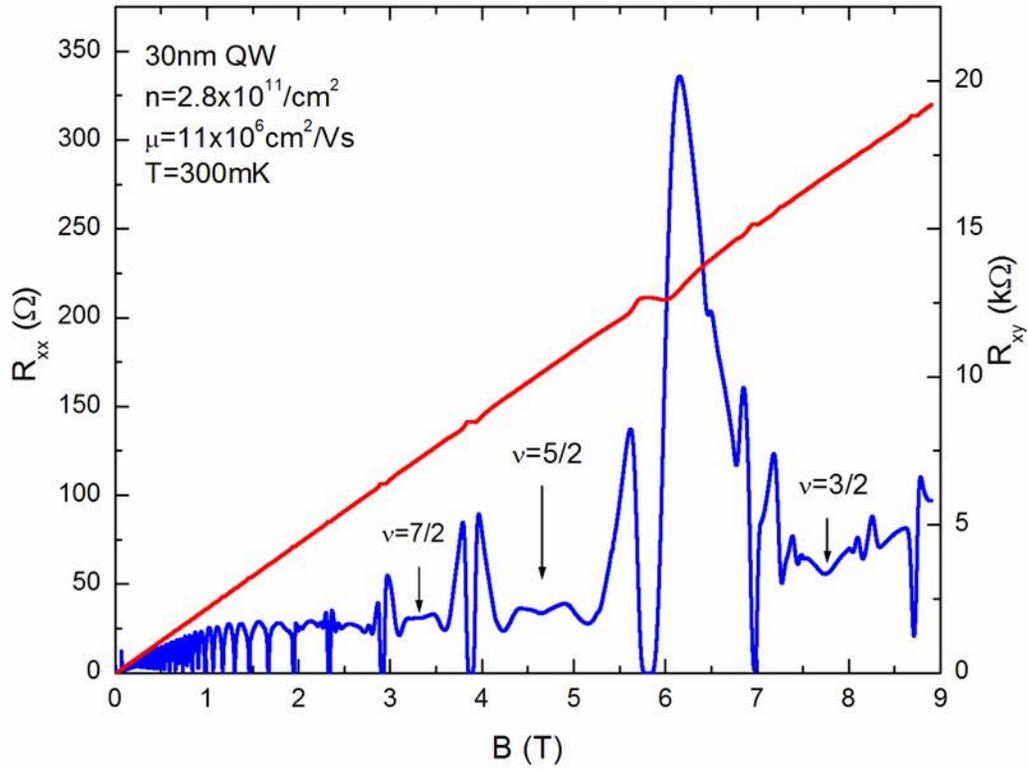

Figure 11: T=0.3K magnetotransport from a relatively low mobility ($11\times10^6$cm$^2$/Vs) doping-well sample. Despite the low mobility and high measurement temperature, many nascent features in the 2$^{nd}$ and higher Landau levels are already visible. The measured resistivity at ν=5/2 is 40.8 Ω/□. Note the similarity of the transport data to that in Figure 8 despite the large difference in mobility.

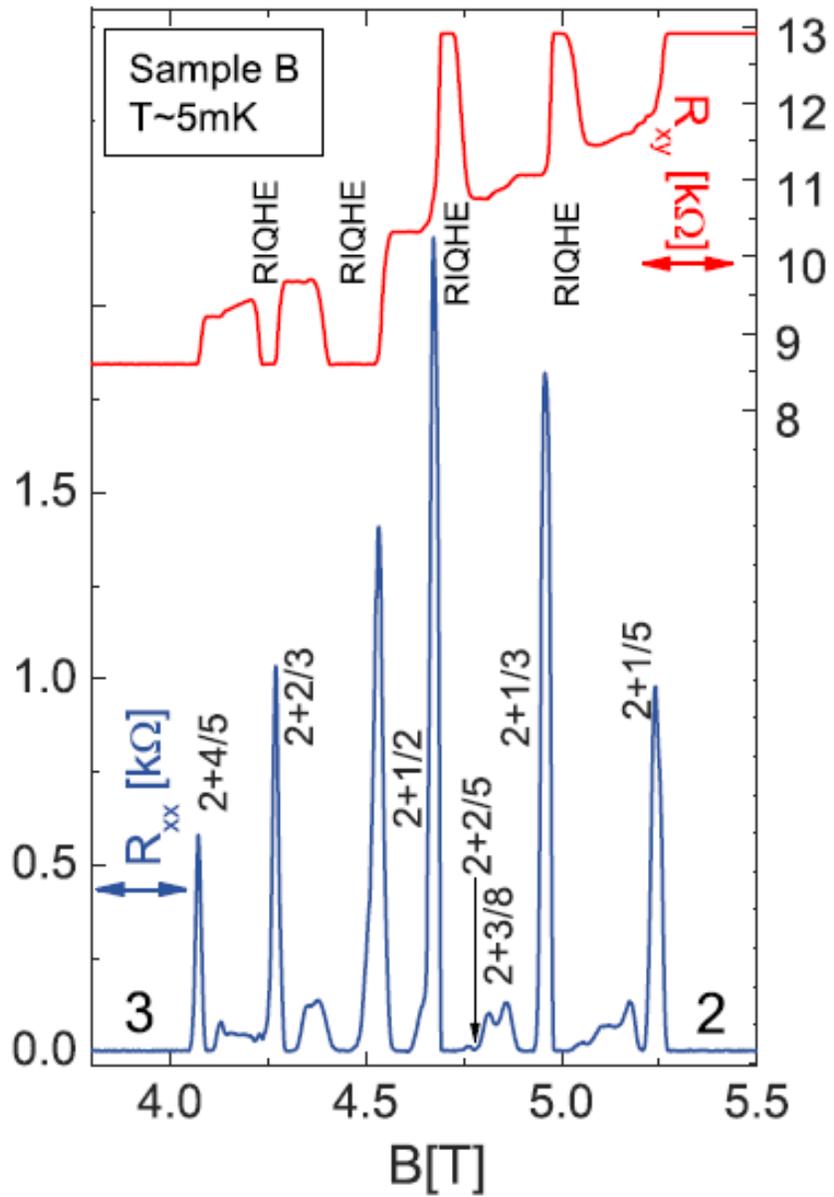

Figure 12: Magnetotransport from an $11 \times 10^6 cm^2/Vs$ mobility sample at T~5mK. We mark the filling factors ν of the observed FQHE and the reentrant integer quantum Hall states. (After Ref. [76]) This is the same sample shown in Figure 11 at T=300mK.

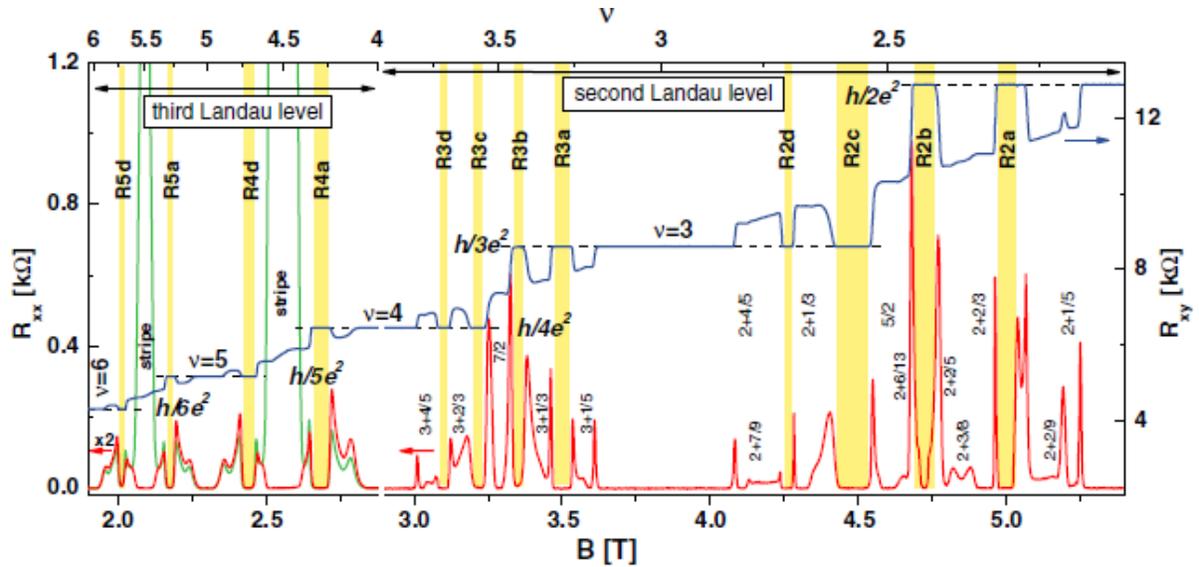

Figure 13: Magnetoresistance in the 2nd (2<ν<4) and the third (4<ν<6) Landau levels. Both the longitudinal ($R_{xx}$) and the Hall ($R_{xy}$) resistances are measured at 77mK in the third Landau level and at 6.9mK in the 2nd Landau level. Reentrant integer quantum Hall states are marked by shaded stripes and the FQHE states by their filling factors. In the third Landau level the two $R_{xx}$ traces shown are measured along mutually perpendicular directions and, for clarity, are magnified by a factor of 2. (After Ref. [77])

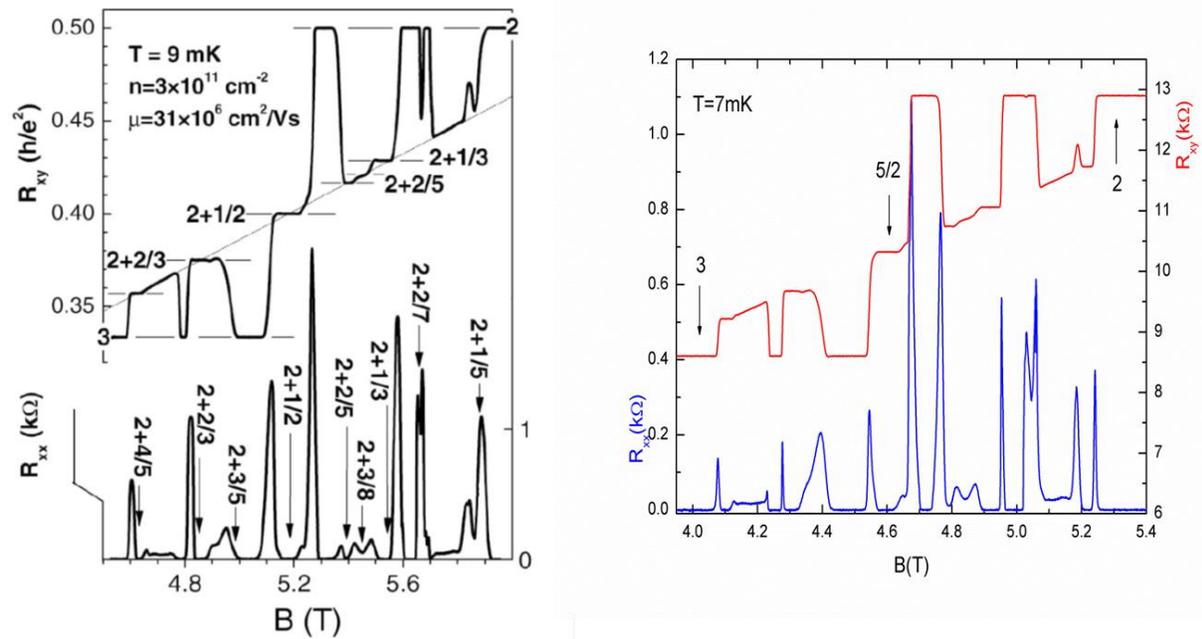

Figure 14: Comparison of data in the 2$^{nd}$ Landau level. On the left is the data of Xia [69] and on the right is the data of Deng [75] with $n_e$=2.8x10$^{11}$cm$^{-2}$ and $\mu$=15x10$^6$cm$^2$/Vs.